\documentclass[lettersize,journal]{IEEEtran}
\usepackage{amsmath,amsfonts}
\usepackage{algorithmic}
\usepackage{algorithm}
\usepackage{array}
\usepackage[caption=false,font=normalsize,labelfont=sf,textfont=sf]{subfig}
\usepackage{textcomp}
\usepackage{stfloats}
\usepackage{url}
\usepackage{verbatim}
\usepackage{graphicx}
\usepackage{cite}
\usepackage{bm}
\usepackage{amssymb}
\usepackage{booktabs}

\AtBeginDocument{%
  \setlength{\abovedisplayskip}{4pt plus 1pt minus 2pt}%
  \setlength{\belowdisplayskip}{4pt plus 1pt minus 2pt}%
  \setlength{\abovedisplayshortskip}{2pt plus 1pt minus 1pt}%
  \setlength{\belowdisplayshortskip}{2pt plus 1pt minus 1pt}%
}

\begin{document}
\title{BDFlow-3DRM: Height-Coherent 3D Radio Map Construction via Bi-Dynamical Flow Matching}

\author{
    Jun Yu,
    Meixia Tao,~\IEEEmembership{Fellow,~IEEE,}
    Shu Sun,~\IEEEmembership{Senior Member,~IEEE}
    \thanks{Part of this work was submitted to IEEE GLOBECOM 2026. The authors are with the School of Information Science and Electronic Engineering, Shanghai Jiao Tong University, Shanghai, 200240, China (emails: \{shjdyujun, mxtao, shusun\}@sjtu.edu.cn).}
}



\maketitle

\begin{abstract}
Three-dimensional (3D) radio map (RM) is a key enabler for environment-aware communications in low-altitude wireless scenarios by providing site-specific channel priors indexed by spatial locations. However, existing 3D RM construction methods lack effective modeling of the height dimension, which limits their generalization to unseen height configurations and degrades construction coherence across height layers. In this paper, we propose BDFlow-3DRM, a bi-dynamical flow matching framework for 3D RM construction. Specifically, the 3D RM construction problem is formulated as a deterministic probability flow in a semantic latent space. BDFlow-3DRM learns continuous height-aware representations from flexible transceiver height inputs, enhancing geometric awareness. Meanwhile, its bi-dynamical design explicitly models bidirectional dependencies across neighboring height layers, so that RMs at different height layers can be constructed jointly. 
Extensive experiments on multiple datasets, covering diverse simplified and realistic urban scenarios, validate the effectiveness of BDFlow-3DRM. Compared with diffusion-based baselines, it reduces the normalized mean square error (NMSE) by 28.6\% and attains a 180-fold reduction in inference complexity. 
More importantly, with only 20 training receiver-height layers, BDFlow-3DRM maintains accurate prediction over a wide continuous receiver-height range from 1 to 120\,m under variable transmitter heights, highlighting its practical potential for large-scale 3D RM construction.
\end{abstract}

\begin{IEEEkeywords}
3D radio map, height-aware modeling, low-altitude intelligent networks, flow matching, zero-shot generalization.
\end{IEEEkeywords}

\section{Introduction}

\IEEEPARstart{T}{he} rapid rise of the low-altitude economy and 6G space-air-ground integrated networks is transforming wireless communications from conventional terrestrial connectivity to pervasive three-dimensional (3D) coverage. Emerging applications, including unmanned aerial vehicle (UAV) swarms, aerial logistics, urban air mobility, and low-altitude intelligent transportation, require wireless networks to support highly dynamic nodes distributed throughout large-scale 3D airspaces while maintaining reliable and ubiquitous connectivity \cite{bi2019engineering,zeng2019accessing}. In such scenarios, wireless propagation exhibits complex spatial propagation patterns due to building blockages and intricate interactions between electromagnetic waves and the surrounding environments. Consequently, future low-altitude intelligent networks increasingly rely on accurate environmental awareness and 3D channel characterization to support critical functionalities such as beam management, radio resource allocation and trajectory planning.

Radio maps (RMs) \cite{romero2022radio} have emerged as a promising solution for enabling these capabilities in future wireless systems. By establishing mappings between spatial locations and wireless propagation characteristics, RM facilitates proactive acquisition of site-specific channel information without exhaustive online measurements, thereby supporting intelligent decision-making and improving overall network efficiency. Closely related representations include channel knowledge map (CKM) \cite{9373011} and channel fingerprint (CF) \cite{ref_MML}. Although these representations differ in the specific channel metrics considered, they share the common objective of characterizing site-specific wireless propagation properties. Specifically, RM primarily focuses on large-scale received signal strength distributions, while CKM and CF may provide richer channel information, including propagation delay, angle-of-arrival (AoA), angle-of-departure (AoD), and fading statistics. The goal of this paper is to investigate 3D RM construction for low-altitude intelligent networks.

Early RM construction approaches mainly rely on classical interpolation techniques or ray tracing (RT) \cite{he2018design}. Classical interpolation approaches, such as Kriging, K-nearest neighbors and matrix completion, estimate unknown regions from sparse measurements. These approaches, however, generally neglect environmental topology and require dense measurements to achieve satisfactory performance. In contrast, RT explicitly simulates electromagnetic propagation based on environmental geometries and material properties. Nevertheless, the substantial computational burden and the dependence on detailed 3D scene information significantly limit the practical applicability of RT in large-scale and latency-sensitive deployment scenarios.

Recently, deep learning has revolutionized RM construction, achieving a favorable balance between accuracy and computational efficiency. Existing studies initially focus on two-dimensional (2D) RM construction, where radio propagation is represented on discretized planar grids and formulated as image reconstruction or generation tasks. Within this paradigm, convolutional neural network (CNN)-based approaches, such as RadioUNet \cite{ref_RadioUNet}, employ encoder--decoder architectures for efficient pathloss prediction. The generative adversarial networks (GANs), including RME-GAN \cite{ref_RME_GAN}, further improve modeling capacity by capturing complex multipath propagation effects through adversarial learning. More recently, diffusion-based frameworks, such as RadioDiff \cite{ref_RadioDiff} and CKMDiff \cite{ref_CKMDiff}, have demonstrated impressive generation quality through iterative denoising. Flow matching-based approaches represented by RadioFlow \cite{jia2025radioflow} further improve generation efficiency by learning the probability flow.
However, the aforementioned methods fundamentally rely on planar assumptions and are primarily designed for fixed-height communication scenarios (e.g., 1.5\,m). They become insufficient for low-altitude applications, where both transmitter (Tx) and receiver (Rx) exhibit height variations. This limitation has motivated increasing research interest in 3D RM construction.

Recent advances in 3D RM construction have led to a variety of learning paradigms. CNN-based models learn direct mappings from input conditions to 3D RMs. Representative studies include UniRM \cite{ref_UniRM}, which integrates a U-Net backbone with prompt-based conditioning to improve cross-environment generalization, and DF-3DRME \cite{DF3DRME}, which employs a 3D U-Net
architecture combined with a two-stage coarse-to-fine refinement strategy for sampling-free 3D RM generation. Autoencoder-based methods provide an alternative reconstruction paradigm by learning compact latent representations from sparse observations. For instance, 3D convolutional autoencoders combined with UAV-assisted measurements \cite{ref_ActiveTraj} enable efficient RM reconstruction under limited observation conditions. GAN-based approaches have also been extended to 3D RM construction. For example, 3D-DCRGAN \cite{ref_3D_DCRGAN} utilizes adversarial learning for reconstruction under constrained UAV observations, while RadioGen3D \cite{ref_RadioGen3D} employs a conditional GAN trained on large-scale synthesized datasets to generate 3D RM.
Following the success in 2D RM modeling, diffusion-based approaches have also been adapted to 3D scenarios. Specifically, 3DCKMDiff \cite{ref_3DCKMDiff} formulates RM construction as a diffusion-based inverse problem under measurement supervision. 3D-RadioDiff \cite{ref_3DRadioDiff} performs height-conditioned slice-wise generation based on 2D diffusion models, while RadioDiff-3D \cite{ref_RadioDiff3D} adopts a full 3D diffusion framework. RadioLAM \cite{ref_RadioLAM} improves generalization through measured-data augmentation and a mixture-of-experts learning architecture. 
Transformer-based methods provide a sequential generation paradigm for 3D RM construction. PILOT \cite{PILOT} employs a physics-guided autoregressive transformer to characterize wavefront propagation, whereas FARM \cite{FARM} combines transformer encoders with diffusion decoders to exploit both sequence modeling and generative capabilities. 
In parallel, neural rendering approaches provide an alternative perspective by explicitly representing continuous radio fields. BiWGS \cite{ref_BiWGS} renders wireless radiation fields with strong physical interpretability. However, such neural rendering methods generally exhibit substantial dependence on scene-specific measurements, which limits scalability and transferability across diverse environments.

Despite the aforementioned progress, existing learning-based methods for 3D RM construction still lack effective modeling of the height dimension. This limitation is particularly critical in low-altitude wireless scenarios, where Tx and Rx heights vary over a wide range and the corresponding signal distributions may evolve along the height dimension due to changing blockage conditions. Many existing methods model the height dimension in an overly simplified manner, for example by treating different height layers as independently generated slices or by using predefined height discretization. As a result, they are often restricted to a finite set of height configurations and may produce cross-height constructions with weakened structural coherence. Voxel-based approaches can partially capture cross-height correlations, yet their representation capability remains limited by voxel resolution, making accurate 3D RM modeling computationally demanding under fine-grained discretization. Consequently, how to achieve effective modeling of the height dimension for accurate 3D RM construction remains an open challenge.

To address this challenge, this paper proposes BDFlow-3DRM, a bi-dynamical flow matching framework for 3D RM construction. The 3D RM construction problem is formulated as a deterministic probability flow in a semantic latent space, where the RMs at different height layers are constructed jointly. Unlike diffusion models that rely on stochastic denoising, flow matching learns a deterministic trajectory from noise to target representations. This mathematical property not only accelerates inference but also makes it uniquely suitable for explicitly injecting correlations across height layers. The main features and novelties of our BDFlow-3DRM are two-fold. 

First, BDFlow-3DRM enables zero-shot inference at unseen height configurations. Instead of treating 3D RM construction as the fitting of discrete height slices, the proposed framework learns continuous height-aware representations from flexible Tx/Rx height inputs. This is achieved through a dynamic height sampling and dual height encoding strategy. During training, the model is conditioned on varied height configurations, and jointly encodes the absolute Rx height and the relative Tx height, thereby enhancing geometric awareness.

Second, BDFlow-3DRM improves construction accuracy and cross-height construction coherence by explicitly exploiting dependencies across neighboring height layers. Specifically, we develop a bi-dynamical inter-layer correlation mechanism, where bi-dynamical refers to the bidirectional modeling of lower- and upper-layer dependencies during the flow construction process. This design enables the RM at each target height layer to be jointly constructed with its neighboring layers through adaptive bidirectional fusion. A ranking loss is further introduced to regularize the learned dependency strengths and promote more reliable dependency modeling.

%太长了，写成两句话
To overcome the height coverage limitation of the existing dataset UrbanRadio3D\cite{ref_RadioDiff3D} (up to 20 meters), we generate the low-altitude intelligent network Radio3D (LAIN-Radio3D) dataset\footnote{The dataset is available at: \url{https://github.com/sjtuyj25-oss/LAIN-Radio3D}}, a large-scale dataset tailored for low-altitude scenarios. The dataset provides extensive 3D height coverage with Rx heights up to 120\,m, variable Tx heights from 7\,m to 77\,m, and 5.07 million RM slices.

Extensive experiments on UrbanRadio3D, LAIN-Radio3D, and LAMBDA\cite{lambda2026} datasets, covering diverse simplified and realistic urban scenarios, validate the effectiveness of BDFlow-3DRM. Compared with diffusion-based baselines, it reduces the normalized mean square error (NMSE) by 28.6\% and attains a 180-fold reduction in inference complexity. More importantly, with only 20 training Rx-height layers, BDFlow-3DRM maintains accurate prediction over a wide continuous Rx-height range from 1 to 120\,m under variable Tx heights. It also exhibits robust zero-shot and few-shot transferability to realistic urban environments, highlighting its practical potential for large-scale 3D RM construction in complex low-altitude scenarios.

The remainder of this paper is organized as follows. Section II provides the preliminaries on generative flow matching and formulates the 3D RM construction problem. Section III details the proposed BDFlow-3DRM. Section IV presents the experimental evaluations, and Section V concludes the paper.

\section{Preliminaries and Problem Formulation}

\subsection{Flow Matching Generative Model}
Diffusion models have emerged as a stable and high-quality alternative to GANs and variational autoencoders. DDPM \cite{ho2020denoising} generates data through a forward noising process and a learned reverse denoising process. However, its reliance on hundreds of denoising steps makes sampling computationally expensive. Recently, flow matching has attracted growing attention as a more efficient generative paradigm\cite{lipman2023flowmatchinggenerativemodeling}. It directly models the probability flow ordinary differential equation (ODE) that defines a continuous-time transformation from a simple prior distribution (e.g., standard Gaussian) to the complex data distribution $p_{\text{data}}$. Formally, let $\mathbf{x}_0 \sim p_{\text{data}}$ be a ground-truth data sample and $\bm{\epsilon} \sim \mathcal{N}(\mathbf{0}, \mathbf{I})$ be a sample from the standard Gaussian prior. A conditional flow $\mathbf{x}_t$ from the data point $\mathbf{x}_0$ towards the noise sample $\bm{\epsilon}$ is constructed as a linear interpolation over a time variable $t \in [0, 1]$, given by
\begin{equation}
    \mathbf{x}_t = (1 - t) \mathbf{x}_0 + t \bm{\epsilon}.
    \label{eq:conditional_flow}
\end{equation}
The derivative of this path defines the conditional velocity field
$\mathbf{v}_t \triangleq \frac{\mathrm{d}\mathbf{x}_t}{\mathrm{d}t}
= \bm{\epsilon} - \mathbf{x}_0$,
which points from the data point towards its corresponding noise sample. Since directly regressing the marginal velocity field is intractable, flow matching trains a neural network (NN) $\mathbf{v}_{\theta}(\mathbf{x}_t, t)$ to approximate the conditional velocity field $\mathbf{v}_t$ via the regression loss,
\begin{equation}
    \mathcal{L}_{\text{FM}}(\theta)
    =
    \mathbb{E}_{t, \mathbf{x}_0, \bm{\epsilon}}
    \left\|
    \mathbf{v}_{\theta}(\mathbf{x}_t, t)
    -
    (\bm{\epsilon}-\mathbf{x}_0)
    \right\|^2 .
    \label{eq:flow_matching_loss}
\end{equation}

After training, the learned velocity field $\mathbf{v}_{\theta}(\mathbf{x}_t,t)$ approximates the target conditional velocity field $\mathbf{v}_t$. During inference, samples are generated by solving the probability flow ODE
$\frac{\mathrm{d}\mathbf{x}_t}{\mathrm{d}t}
=
\mathbf{v}_{\theta}(\mathbf{x}_t,t)$
from $t=1$ back to $t=0$. The deterministic nature of this ODE supports efficient numerical solvers, thus enabling the generation of high-quality samples with significantly reduced computational overhead compared to stochastic reverse processes in diffusion models.

\subsection{Problem Formulation of 3D RM Construction}
The goal of 3D RM construction in our work is to estimate the spatial distribution of a channel metric (e.g., received signal strength) in a volumetric region solely based on environmental geometry and Tx configurations. We define a target 3D RM as a multi-layer tensor $\mathcal{R}\in\mathbb{R}^{L \times W \times D}$, where $D$ denotes the number of height layers, and $L \times W$  represents the number of uniformly distributed grids in the horizontal plane. Each element $\mathcal{R}_{l,w,d}$ indicates the received signal strength at the coordinate associated with the horizontal grid $(l,w)$ at the $d$-th height layer. The environmental information is defined by a 3D occupancy grid $\mathcal{E} \in \{0,1\}^{L \times W \times H_{\text{env}}}$, where $H_{\text{env}}$ denotes the vertical grid size of the environment and $\mathcal{E}_{x_i, y_i, h_i} = 1$ indicates the presence of occlusions at the 3D coordinate $(x_i, y_i, h_i)$, characterizing the distribution of occlusions in 3D space. Note that $H_{\text{env}}$ is determined by the fixed spatial resolution of the 3D environmental topology, whereas $D$ represents the number of target Rx height layers, which can be flexibly sampled regardless of the environmental grid size. The Tx information includes its horizontal location ($x_{tx},y_{tx}$) and height $h_{tx}$, denoted as $\mathcal{T}$.

The objective is to train an NN $\Phi_\theta(\cdot)$ with parameters $\theta$ to predict the 3D RM $\mathcal{R}$ of the target region based on environmental features $\mathcal{E}$, Tx information $\mathcal{T}$, and a vector of target heights $\mathcal{H} \in \mathbb{R}^{1 \times D}$. This is a conditional generation problem, where the model is trained to minimize the discrepancy between the predicted RM $\hat{\mathcal{R}} \in \mathbb{R}^{L \times W \times D}$ and the ground truth $\mathcal{R}$. This problem can be formulated as

\begin{equation}
    \begin{aligned}
        \min_{\theta} &\ \mathcal{L}\left(\hat{\mathcal{R}},\mathcal{R}\right), \\
        \text{s.t.} &\ \hat{\mathcal{R}} = \Phi_\theta \left(\mathcal{E}, \mathcal{T}, \mathcal{H}\right),
    \end{aligned}
    \label{eq:1}
\end{equation}
where $\mathcal{L} (\cdot)$ is a suitable loss function measuring the difference between the two 3D RMs. 

\begin{figure*}[!t]
    \centering
    \includegraphics[width=0.86\linewidth]{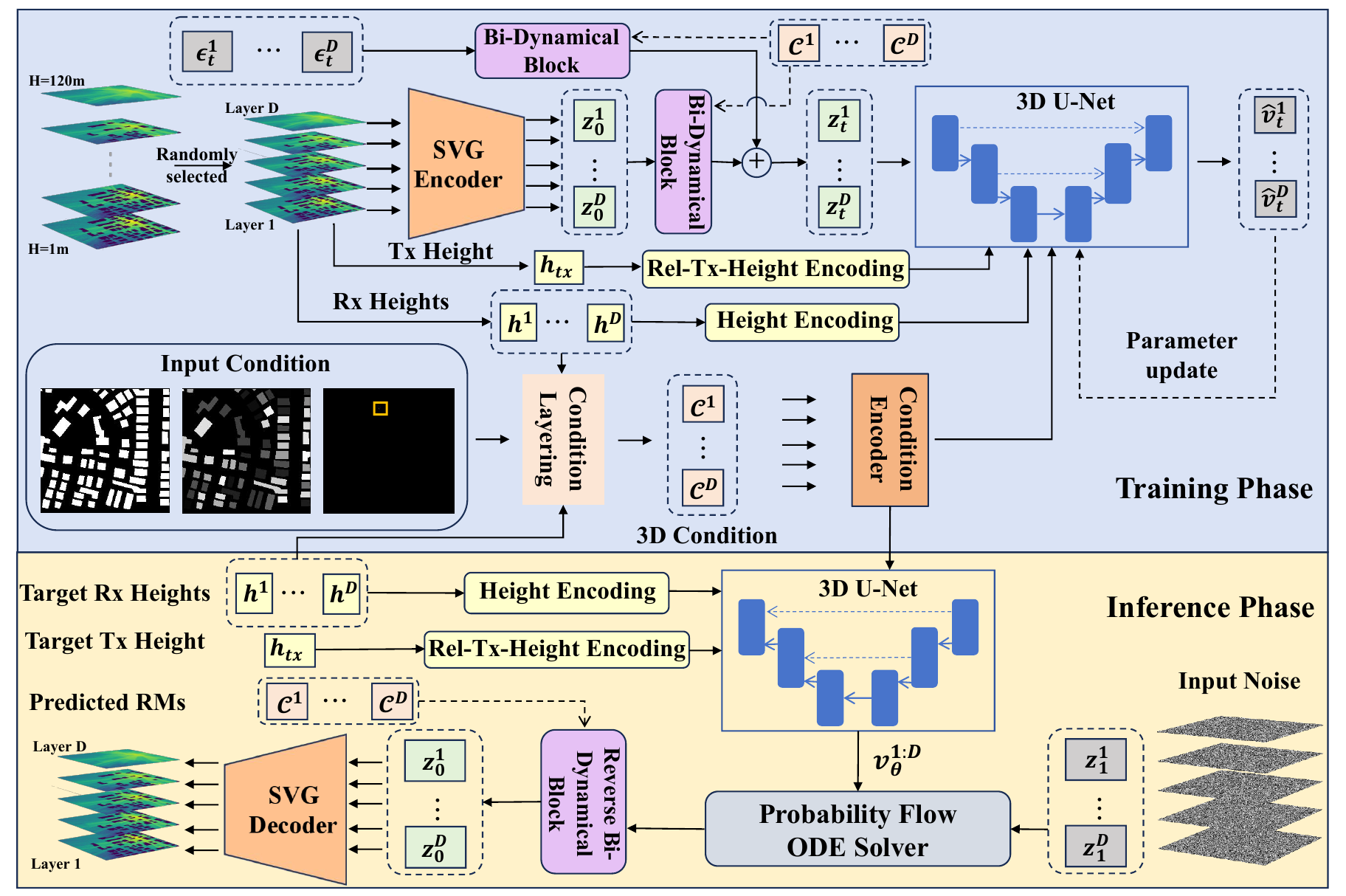}
    \caption{Overview of the proposed BDFlow-3DRM.}
    \label{Flowchart_all}
\end{figure*}

\section{The Proposed Method}

The overall architecture of BDFlow-3DRM is illustrated in Fig.~\ref{Flowchart_all}. The framework formulates 3D RM construction as a deterministic probability flow process in semantic latent space. Based on the environmental information, Tx configuration, and target Rx heights, BDFlow-3DRM generates the complete 3D RM through continuous height-aware representations and adaptive inter-layer dependency modeling. It consists of training and inference phases.

During training, a dynamic height sampling strategy selects a subset of RM layers from the 3D RM datasets, which are then encoded by the self-supervised representations for visual generation (SVG) \cite{shi2026latentdiffusionmodelvariational} encoder into semantic latent representations. Meanwhile, the condition layering module processes the environmental information to produce height-aware 3D conditions, and the dual height encoding module maps Tx and Rx heights into continuous embeddings. The bi-dynamical inter-layer correlation mechanism couples neighboring latent layers and their corresponding noises, and constructs a probability flow path between the fused noises and latent representations. Conditioned on the environment and height embeddings, a 3D U-Net is trained to predict the latent velocity field.

During inference, generation starts from Gaussian noise in latent space. Conditioned on the environment and transceiver height embeddings, a probability flow ODE solver transforms noisy latent variables into fused latent representations. The resulting features are decoupled by the reverse bi-dynamical block and decoded by the SVG decoder into the 3D RM.

The designs of the dynamic height sampling and encoding, SVG-enhanced 3D U-Net backbone architecture and bi-dynamical inter-layer correlation mechanism are detailed in the following subsections.

\subsection{Dynamic Height Sampling and Encoding}
\label{sec:Dynamic Height Sampling and Encoding}
The proposed framework incorporates a dynamic height sampling and dual height encoding strategy. Specifically, instead of processing all $N$ height layers during each training step, where $N$ denotes the total number of discrete height layers available in datasets, the model randomly selects a fixed subset containing $D$ height layers from the complete height pool as the training input. This sampling strategy exposes the model to diverse combinations of height configurations, thereby encouraging the learning of signal distribution characteristics rather than memorizing specific discrete height patterns. Furthermore, this mechanism provides computational flexibility while maintaining sufficient vertical correlation information among sampled layers.

Although dynamic height sampling exposes the model to diverse height combinations, accurate 3D RM construction still requires explicit awareness of the Tx--Rx height geometry. In complex low-altitude environments, height variation affects both the absolute Rx position and its relative geometric relationship to the Tx, which jointly shape the signal distribution. Therefore, a dual height encoding mechanism is introduced to jointly model the absolute Rx height and the relative height difference between the Tx and Rx. For the $d$-th sampled layer, its absolute height is given by $h^{d}$, and we define the Tx-relative height as
\begin{equation}
\delta h_{tx}^{d} = h_{tx} - h^{d}.
\end{equation}
To represent continuous height information, this module maps scalar height values into a continuous high-dimensional embedding space. Given an even embedding dimension $d_e$, we employ the sinusoidal mapping $\text{PE}(\cdot): \mathbb{R} \rightarrow \mathbb{R}^{d_e}$, defined as
\begin{equation}
\begin{aligned}
\text{PE}(u)_{2m} &= \sin\!\left(\frac{u}{10000^{\frac{2m}{d_e}}}\right),\\
\text{PE}(u)_{2m+1} &= \cos\!\left(\frac{u}{10000^{\frac{2m}{d_e}}}\right),
\end{aligned}
\end{equation}
where $m = 0, 1, \dots, \frac{d_e}{2}-1$ denotes the embedding index. The exponentially scaled frequencies generated by the term
$10000^{\frac{2m}{d_e}}$ enable the embedding to simultaneously capture local and global variations associated with different height levels. Accordingly, the absolute Rx height embedding and the Tx-relative height embedding of the $d$-th sampled layer are expressed as $\mathbf{e}_{rx}^{d} = \text{PE}\!\left(h^{d}\right)$ and $\mathbf{e}_{tx}^{d} = \text{PE}\!\left(\delta h_{tx}^{d}\right)$, respectively. These two representations are then separately injected into the corresponding layer-wise branch of the network, ensuring that each height layer is explicitly conditioned on the current physical height.

Crucially, by exploiting the continuous characteristics of the sinusoidal embedding space, the proposed representation enables the model to establish a smooth mapping between physical height variations and complex electromagnetic propagation features. Such a structured height representation facilitates interpolation and extrapolation across unseen heights, enabling accurate inference for unseen height layers both within and beyond the training height range. Consequently, the network learns height-aware representations of signal distribution instead of merely fitting fixed numerical patterns associated with discrete height indices.

\subsection{SVG-Enhanced 3D U-Net Architecture}

BDFlow-3DRM employs an SVG-enhanced 3D U-Net architecture as the core generative backbone for conditional 3D RM generation. The entire architecture is established upon a semantic latent space constructed by the SVG framework, which facilitates efficient generation while preserving high construction fidelity. Specifically, SVG adopts a dual-branch encoder architecture composed of a pre-trained feature extractor $\mathcal{F}_{feat}$ and a lightweight learnable residual encoder $\mathcal{F}_{res}$. The feature extractor $\mathcal{F}_{feat}$ utilizes a frozen self-supervised vision foundation model (e.g., DINOv3 \cite{siméoni2025dinov3}) to extract semantic representations $\mathbf{f}_{sem}$, capturing global physical contexts and signal distributions. In parallel, $\mathcal{F}_{res}$ is dedicated to capturing fine-grained spatial details, producing complementary residual features $\mathbf{f}_{res}$. These two representations are concatenated to construct the latent representation $z = \text{Concat}(\mathbf{f}_{sem}, \mathbf{f}_{res})$, which preserves strong semantic discrimination and detailed structural information. The subsequent flow matching process is performed within this semantic latent space, where $\mathbf{z}_0 \text{ and } \mathbf{z}_t$ denote the latent variables at the initial state and the flow timestep $t$, respectively. A decoder $\mathcal{D}$ is then trained to reconstruct the original input from this latent representation. 

In the 3D RM construction task, the SVG encoder is applied to each height layer to obtain layer-wise latent features, which are then stacked along the vertical dimension to form the 3D latent representation $\mathbf{z}_0^{1:D} = \{\mathbf{z}_0^1, \ldots, \mathbf{z}_0^D\}$. This stacked latent volume serves as the initial state of the flow matching process. The structured feature space provided by SVG establishes a robust foundation for modeling complex electromagnetic propagation characteristics. As a result, the burden of learning the high-dimensional data distribution is significantly reduced, leading to faster optimization convergence and improved generation quality. Based on this, we deploy a 3D U-Net as the backbone for velocity field prediction. Leveraging 3D convolutional modules, the 3D U-Net jointly captures both intra-plane signal distribution patterns and vertical inter-layer propagation correlations during feature extraction. 

For conditional guidance, the flow timestep $t$ is injected into the network through a time embedding module, allowing the model to adaptively adjust the velocity field according to the current state along the probability flow trajectory. Similarly, the height encoding representations generated during the dynamic sampling stage are incorporated through the same embedding injection mechanism. Furthermore, environmental features $\mathcal{E}$ and Tx information $\mathcal{T}$ are first transformed into the building layout map $\mathbf{M}_b$ and height map $\mathbf{M}_h$, and the Tx position map $\mathbf{M}_{tx}$, respectively. Subsequently, we introduce a condition layering step to strengthen the height-dependent perception capability of the model. Specifically, by comparing the height map with the target Rx heights, we extract building distribution slices for each layer, while the layout and Tx maps are shared among all height layers, as depicted in the condition layering module of Fig.~\ref{Flowchart_all}. Consequently, the layer-wise 3D condition $\mathcal{C}^{1:D}=\{\mathcal{C}^1,\dots,\mathcal{C}^D\}$ is constructed. These conditions are first encoded into latent feature sequences by a shared condition encoder $\tau_\psi$ as

\begin{equation}
\hat{\mathcal{C}}^{1:D} = \tau_\psi(\mathcal{C}^{1:D}),
\end{equation}
and then injected as guidance information into various levels of the 3D U-Net through the cross-attention mechanism, spatially constraining the received signal strength distributions with high precision.

\subsection{Bi-Dynamical Block}

\begin{figure*}[!t]
    \centering
    \includegraphics[width=0.85\linewidth]{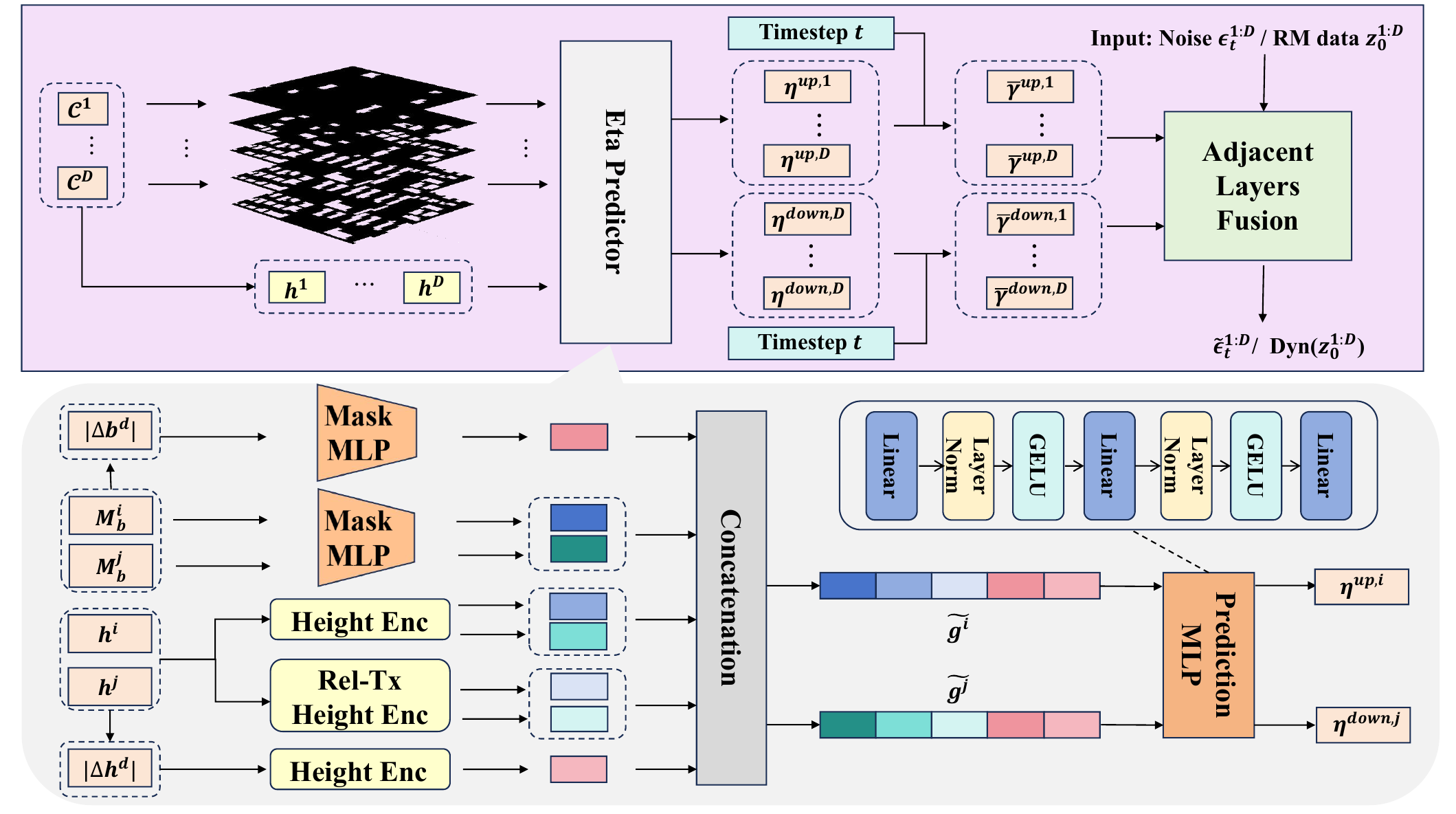}
    \caption{Network architecture for the Bi-Dynamical Block}
    \label{Flowchart_layer}
\end{figure*}

Although signal distributions vary across different heights, the propagation behavior is fundamentally governed by the same physical environment and therefore naturally exhibits vertical correlations. The proposed bi-dynamical block, whose architecture is illustrated in Fig.~\ref{Flowchart_layer}, is designed to explicitly incorporate the interactions among RMs at neighboring height layers during the probability flow process, thereby effectively capturing and exploiting vertical dependencies. Furthermore, the designed mixing mechanism is conditioned on the flow timestep $t$, enabling the model to simultaneously perceive both temporal evolution along the flow trajectory and vertical variations across height layers. This dual height-time awareness facilitates more effective modeling of cross-layer dependencies. Based on this design, the mathematical formulations of the proposed bi-dynamical block in the probability path formulation and generative process are presented as follows.
\subsubsection{Probability Path Formulation Process}
\label{sec:Probability Path Formulation}
Inspired by dynamical diffusion \cite{guo2025dynamicaldiffusionlearningtemporal}, which introduces historical state interactions into the diffusion trajectory, the noisy state ${{\mathbf{x}}_t^s}$ is formulated as
 \begin{equation} \label{eq:state_combination} 
\mathbf{x}_{t}^{s} = \sqrt{\bar{\gamma}_{t}} \cdot \left( (1-t)\, \mathbf{x}_{0}^{s} + t \, \bm{\epsilon}_{t}^{s} \right) + \sqrt{1 - \bar{\gamma}_{t}} \cdot \mathbf{x}_{t}^{s-1},
\end{equation}
where ${{\bar{\gamma}}_t}$ serves as a schedule parameter controlling the dependency on the previous state. However, dynamical diffusion is primarily developed for sequential data, where the inherent temporal ordering restricts the interaction to unidirectional dependencies from historical states to the current state. In contrast, height layers in the 3D RM coexist simultaneously within the spatial domain and exhibit mutual correlations among neighboring layers. This characteristic allows 3D RM construction to jointly exploit correlations from both lower and upper neighboring layers. Motivated by this observation, we extend the time-dependent dependency control in dynamical diffusion to a bidirectional form by introducing two height-related coefficients, $\bar{\gamma}_{t}^{\text{down},d}$ and $\bar{\gamma}_{t}^{\text{up},d}$, for the $d$-th target height layer, which regulate the lower and upper layer dependencies, respectively. Their concrete parameterization will be presented in Section~\ref{sec:Height-related}. The proposed bi-dynamical block then models the latent representation at each height through a bidirectional fusion process that recursively incorporates these dependencies across neighboring height layers. To provide a compact explicit form of this bidirectional fusion, we first introduce a normalization factor $K_t^d$ to govern the overall weight distribution, given by
\begin{equation}
    K_{t}^{d} = 1 + \sqrt{\frac{(1-\bar{\gamma}_{t}^{\text{down},d})\bar{\gamma}_{t}^{\text{up},d}(1-\bar{\gamma}_{t}^{\text{up},d-1})}{\bar{\gamma}_{t}^{\text{up},d-1}}}.
    \label{eq:Normalization_Factor}
\end{equation}
With this normalization, the fused latent representation for the $d$-th height layer is expressed as
\begin{equation}
\begin{split}
    \mathbf{z}_{t}^{d} = \frac{1}{K_{t}^{d}} \Bigg[ & \sqrt{\bar{\gamma}_{t}^{\text{up},d} \bar{\gamma}_{t}^{\text{down},d}} \left( (1-t) \mathbf{z}_{0}^{d} + t \bm{\epsilon}_{t}^{d} \right) \\
&\hspace{-2.5em}+ \sqrt{1-\bar{\gamma}_{t}^{\text{up},d}} \mathbf{z}_{t}^{d+1} + \sqrt{\frac{(1-\bar{\gamma}_{t}^{\text{down},d})\bar{\gamma}_{t}^{\text{up},d}}{\bar{\gamma}_{t}^{\text{up},d-1}}} \mathbf{z}_{t}^{d-1} \Bigg],
\end{split}
\label{eq:Layer_Correlation}
\end{equation}
where $\mathbf{z}_{t}^{d-1}$ and $\mathbf{z}_{t}^{d+1}$ denote the fused latent representations of the lower and upper neighboring height layers, and $\bar{\gamma}_{t}^{\text{up},d}$ and $\bar{\gamma}_{t}^{\text{down},d}$ denote the corresponding upward and downward dependency coefficients. To more clearly present the recursive construction of the bi-dynamical process, we decompose it into two stages: lower-layer recursive fusion and upper-layer recursive fusion. We first formulate Mid-Dyn and its corresponding correlated noise $\widetilde{\bm{\epsilon}}_t^{mid,d}$ through the lower-layer recursive fusion as
\begin{equation}
\begin{split}
\text{Mid-Dyn}\left(\mathbf{z}_0^d, \bar{\gamma}_{t}^{\text{down},d}\right) = 
\sqrt{\bar{\gamma}_{t}^{\text{down},d}} \cdot \mathbf{z}_0^d  + \\
\sqrt{1 - \bar{\gamma}_{t}^{\text{down},d}} \cdot 
\text{Mid-Dyn}\left(\mathbf{z}_0^{d-1}, \bar{\gamma}_{t}^{\text{down},d-1}\right),
\label{eq:midcontext}
\end{split}
\end{equation}

\begin{equation}
\widetilde{\bm{\epsilon}}_t^{mid,d} = 
\sqrt{\bar{\gamma}_{t}^{\text{down},d}} \cdot \bm{\epsilon}_t^d + 
\sqrt{1 - \bar{\gamma}_{t}^{\text{down},d}} \cdot \widetilde{\bm{\epsilon}}_t^{mid,d-1}.
\label{eq:epsilon_recursive_mid}
\end{equation}
Both processes are regulated by the parameter $\bar{\gamma}_{t}^{\text{down},d}$. Based on the lower-layer fusion, we then extend this formulation by introducing the dependency on the upper layers as

\begin{equation}
\begin{split}
\text{Dyn}\left(\mathbf{z}_0^d, \bar{\gamma}_{t}^{\text{up},d}\right) = 
\sqrt{\bar{\gamma}_{t}^{\text{up},d}} \cdot \text{Mid-Dyn}\left(\mathbf{z}_0^d, \bar{\gamma}_{t}^{\text{down},d}\right)   \\
+ \sqrt{1 - \bar{\gamma}_{t}^{\text{up},d}} \cdot 
\text{Dyn}\left(\mathbf{z}_0^{d+1}, \bar{\gamma}_{t}^{\text{up},d+1}\right),
\label{eq:context}
\end{split}
\end{equation}

\begin{equation}
\widetilde{\bm{\epsilon}}_t^d = 
\sqrt{\bar{\gamma}_{t}^{\text{up},d}} \cdot \widetilde{\bm{\epsilon}}_t^{mid,d} + 
\sqrt{1 - \bar{\gamma}_{t}^{\text{up},d}} \cdot \widetilde{\bm{\epsilon}}_t^{d+1}.
\label{eq:epsilon_recursive}
\end{equation}
For boundary layers $d=1$ and $d=D$, the recursion is terminated by assigning zero values to the unavailable neighboring dependency terms.

Collectively, Eqs. \eqref{eq:midcontext} - \eqref{eq:epsilon_recursive} establish a hierarchical bi-directional fusion framework. Within this framework, the Dyn term integrates latent representations from all other height layers, while the noise term $\widetilde{\bm{\epsilon}}_t^{d}$ follows an analogous recursive fusion process. Consequently, the originally independent Gaussian noise samples across different height layers become vertically correlated. Using these recursively defined fusion operators, the fused latent representation in \eqref{eq:Layer_Correlation} can be rewritten in the following concise form as
\begin{equation}
\mathbf{z}_t^d = (1-t) \cdot \text{Dyn}\left(\mathbf{z}_0^d, \bar{\gamma}_{t}^{\text{up},d}\right) + t \cdot \tilde{\bm{\epsilon}}_t^d.
\label{eq:simple Layer Correlation}
\end{equation}
This reformulated expression preserves the structural form of the conventional flow matching probability path while introducing vertical correlations into both the latent and noise components. Importantly, under a fixed environment condition and flow timestep, the recursive fusion acts as a deterministic linear transformation, such that the transformed noise remains Gaussian with a covariance structure determined by the fusion coefficients. Therefore, the proposed bi-dynamical formulation preserves the tractable probability path construction required by flow matching, allowing the NN to effectively learn the corresponding generation dynamics while incorporating adaptive inter-layer dependencies.

\subsubsection{Generative Process and Objective}

Building upon the flow matching framework introduced in Section II, the generation of the 3D RM is formulated as a deterministic probability flow ODE process. Instead of restating the standard ODE formulation, this section focuses on the learning of the continuous velocity field $\mathbf{v}_\theta(\mathbf{z}_t^{1:D}, t, \mathcal{\hat{C}}, \mathcal{H})$ parameterized by the NN. 

Different from conventional flow matching, the proposed bi-dynamical probability path introduces explicit vertical coupling among neighboring height layers. Consequently, the corresponding target velocity field is reformulated to characterize the generation dynamics under the bi-dynamical fusion state. It should be noted that, for a given sampled flow timestep $t$ during training, the bi-dynamical fusion is deterministic once its corresponding fusion coefficients are fixed. Therefore, the recursive fusion defines a deterministic transformation of the latent representation at that timestep, allowing the target velocity to be expressed as the difference between the fused noise $\tilde{\bm{\epsilon}}_t^d$ and the fused latent representation $\text{Dyn}(\mathbf{z}_0^d, \bar{\gamma}_t^{\text{up},d})$,
\begin{equation}
\mathbf{v}_{\text{target}}^d = \tilde{\bm{\epsilon}}_t^d - \text{Dyn}\left(\mathbf{z}_0^d, \bar{\gamma}_t^{\text{up},d}\right).
\label{eq:v_target}
\end{equation}

Accordingly, the training objective is formulated to minimize the mean squared error between the predicted velocity field and this target, given by
\begin{equation}
\mathcal{L}_{\text{FM}}(\theta) = \mathbb{E}_{t, \mathbf{z}_0^{1:D}, \bm{\epsilon}} \left[ \sum_{d=1}^{D} \left\| \mathbf{v}_\theta(\mathbf{z}_t^d, t, \mathcal{C}, \mathcal{H}) - \mathbf{v}_{\text{target}}^d \right\|^2 \right].
\label{eq:fm_loss}
\end{equation}

By minimizing this objective, the network $\mathbf{v}_{\theta}$ implicitly learns the vertical dependency structure embedded in the bi-dynamical probability path. Consequently, the learned velocity field enables the ODE integration process to generate latent representations that naturally preserve vertical correlations across different height layers, thereby improving the height coherence of the constructed 3D RM.

During inference, the probability flow ODE is solved under piecewise fixed fusion states within each discretized integration interval, where the representation evolves in the fused latent space with preserved cross-layer vertical correlations. When moving from one integration interval to the next, the current fused latent representation is first mapped back to the decoupled latent space through the reverse bi-dynamical block, and is then recursively fused again using the coefficients associated with the new timestep. This procedure ensures a consistent transition of the latent state across different fusion coefficients. After completing the ODE integration, the final fused latent representation is transformed into height-specific latent features via the last reverse bi-dynamical block. Since the proposed recursive fusion operation is mathematically invertible by construction, the reverse transformation can remove the weighted dependencies introduced from neighboring height layers and recover the intrinsic channel characteristics of individual heights. The resulting decoupled latent features are subsequently decoded by the SVG decoder $\mathcal{D}$ to reconstruct the final 3D RM.

\subsubsection{Height-related Coefficients}
\label{sec:Height-related}
As discussed in the Section~\ref{sec:Probability Path Formulation}, the height-related coefficients include $\bar{\gamma}_{t}^{\text{up},d}$ and $\bar{\gamma}_{t}^{\text{down},d}$. For notational convenience, we use $\bar{\gamma}_{t}^{\xi,d}$, with $\xi \in \{\text{up},\text{down}\}$, to uniformly denote either coefficient in this subsection. Their concrete parameterization is given next. Intuitively, when the flow timestep $t$ is small, the latent representation remains close to the original RM and should preserve stronger layer-specific information; when $t$ becomes large, the latent representation is increasingly dominated by noise and can benefit more from interactions with neighboring height layers. Accordingly, we adopt the following time-decaying design
\begin{equation}
\bar{\gamma}_{t}^{\xi,d} = \eta^{\xi,d} (1-t) + (1 - \eta^{\xi,d}) = 1 - \eta^{\xi,d} t,
\label{eq:22}
\end{equation}
where $\eta^{\xi,d} \in [0,1]$ denotes the height-dependent factor. Since building distributions vary across heights in urban environments, the strength of vertical correlation is inherently non-uniform. Therefore, a fixed scalar value is insufficient. To address this, we introduce an adaptive eta predictor to estimate $\eta^{\xi,d}$ based on the environmental structures.

The eta predictor takes the building masks $\mathbf{M}_b^d$ and their corresponding physical heights $h^d$ as inputs to dynamically infer the inter-layer correlation factor. Since the downward case is defined analogously, we take the estimation of $\eta^{\text{up},d}$ for neighboring layers $d$ and $d+1$ as an example. The predictor first characterizes their discrepancy through the building distribution difference $\Delta b^{d,d+1} = \frac{1}{LW} \left\| \mathbf{M}_b^{d+1} - \mathbf{M}_b^d \right\|_1$, which provides a scalar descriptor of the structural variation between the two building layouts. To make it compatible with the other feature representations, $\Delta b^{d,d+1}$ is further mapped into an embedding vector $\mathbf{e}_{\Delta b}^{d,d+1}$ through a lightweight projection layer. The building mask $\mathbf{M}_b^d$ is also projected into a compact feature representation $\mathbf{e}_b^d$. In addition, sinusoidal positional encodings are used for the absolute Rx height and the Tx-relative height, producing the embeddings $\mathbf{e}_{rx}^{d}$ and $\mathbf{e}_{tx}^{d}$ defined in Section~\ref{sec:Dynamic Height Sampling and Encoding}, while the inter-layer height difference $\Delta h^{d,d+1} = |h^{d+1} - h^d|$ is encoded separately as $\text{PE}(\Delta h^{d,d+1})$. These features are then concatenated and fed into a lightweight multi-layer perceptron (MLP) to produce the raw correlation score,
\begin{equation}
\text{logits}^{\text{up},d} = \text{MLP}_{\eta} \Big( \big[ \mathbf{e}_b^d,\, \mathbf{e}_{rx}^{d},\, \mathbf{e}_{tx}^{d},\, \mathbf{e}_{\Delta b}^{d,d+1}, \text{PE}(\Delta h^{d,d+1}) \big] \Big).
\end{equation}

\begin{algorithm}[t]
\caption{Calculation of Ranking Loss}
\label{alg:ranking_loss}
\begin{algorithmic}[1]
\REQUIRE Predicted coefficients $\eta^{\text{up},1:D}$, Layer heights $h^{1:D}$, Building masks $\mathbf{M}_b^{1:D}$
\ENSURE Ranking loss $L_{rank}$

\STATE \textit{Step 1: Compute a discrepancy score for layer pairs}
\FOR{$d = 1$ \textbf{to} $D - 1$}
    \STATE $\Delta h^{d,d+1} \gets |h^{d+1} - h^d|$
    \STATE $\Delta b^{d,d+1} \gets \frac{1}{LW} \|\mathbf{M}_b^{d+1} - \mathbf{M}_b^d\|_1$
    \STATE $\text{score}^{d} \gets \Delta h^{d,d+1} + \Delta b^{d,d+1}$
\ENDFOR

\STATE \textit{Step 2: Quantify ranking violations}
\STATE $L_{sum} \gets 0, \quad N_{v} \gets 0$ 
\FOR{$d = 1$ \textbf{to} $D - 2$}
    \IF{$\text{score}^{d+1} > \text{score}^{d}$}
        \STATE $\delta \gets \max(0, \, \eta^{\text{up},d+1} - \eta^{\text{up},d})$ 
        \STATE $L_{sum} \gets L_{sum} + \delta$
        \STATE $N_{v} \gets N_{v} + 1$
    \ENDIF
\ENDFOR

\STATE \textit{Step 3: Aggregate ranking loss}
\IF{$N_{v} > 0$}
    \STATE $L_{rank} \gets L_{sum} / N_{v}$
\ELSE
    \STATE $L_{rank} \gets 0$
\ENDIF
\STATE \RETURN $L_{rank}$

\end{algorithmic}
\end{algorithm}
To prevent severe factor fluctuations from disrupting the early stages of model training, a learnable residual gating mechanism is introduced to stabilize the output. The final height-dependent factor $\eta^{\text{up},d}$ is formulated as
\begin{equation}
    \eta^{\text{up},d} = \eta_{init} \cdot (1 - \sigma(\lambda)) + \sigma(\text{logits}^{\text{up},d}) \cdot \sigma(\lambda),
\end{equation}
where $\eta_{init} = 0.5$ serves as a stable initial prior, and $\lambda$ is a learnable parameter initialized with a negative value. Here, $\sigma(\cdot)$ denotes the sigmoid function. The term $\sigma(\lambda)$ serves as a gating coefficient that balances the prior and the predictor output, while $\sigma(\text{logits}^{\text{up},d})$ represents the normalized output of the eta predictor. The negative initialization encourages the model to initially rely more on the prior and progressively shifts to learned correlation estimation as the training proceeds. 

Since $\eta^{\xi,d}$ lacks direct ground-truth supervision, we employ a composite loss $\mathcal{L}_{total} = \mathcal{L}_{\text{FM}} + \mathcal{L}_{rank}$ to prevent representation collapse. Here, $\mathcal{L}_{\text{FM}}$ is the flow matching loss defined in (\ref{eq:fm_loss}), and Algorithm~\ref{alg:ranking_loss} presents the detailed computation procedure of the auxiliary ranking loss $\mathcal{L}_{\text{rank}}$, also taking the upward case as an example. Specifically, $\mathcal{L}_{\text{rank}}$ regularizes the predicted correlation weights by discouraging neighboring layer pairs with larger discrepancy scores from receiving stronger coupling than pairs with smaller scores. Ultimately, this progressive gating strategy, combined with the discrepancy-aware ranking regularization, enables a fine-grained adaptive regulation of inter-layer dependencies based on the specific environmental structures.

\section{Experimental Results}
In this section, we comprehensively evaluate the performance of BDFlow-3DRM. Section IV-A first introduces the experimental setup, and Section IV-B details the representative baselines and evaluation metrics. Section IV-C presents the overall performance comparison against the baseline methods. Ablation studies on the key network components are discussed in Section IV-D. Furthermore, Section IV-E and Section IV-F provide in-depth analyses of the model's height coherence and the adaptive correlation mechanism driven by the eta predictor, respectively. Section IV-G evaluates the generalization capability of BDFlow-3DRM across unseen Tx/Rx height configurations and realistic urban environments, and Section IV-H reports the computational efficiency and complexity analysis.

\subsection{Experimental Setup}
\subsubsection{Datasets}In this study, we evaluate the performance of BDFlow-3DRM on three datasets: the UrbanRadio3D dataset, our newly constructed LAIN-Radio3D dataset, and the LAMBDA dataset.

\textbf{UrbanRadio3D}: The UrbanRadio3D dataset is a 3D RM dataset simulated based on urban environments. It contains 701 distinct urban layouts via OpenStreetMap. Each geographical area covers $256 \times 256\text{ m}^2$ with a 1\text{-meter} grid resolution, where the corresponding building heights span from 6.6\text{ m} to 19.8\text{ m}. The corresponding RMs are generated using the dominant path model (DPM) \cite{Wahl2005dominant} in the Altair WinProp suite. However, the simulation parameters are primarily constrained to near-ground scenarios. For example, the Tx height is fixed, and the Rx heights are limited to a range of 1\text{ m} to 20\text{ m} in 1\text{-meter} increments. While valuable for 3D RM construction, these restricted height dimensions limit its applicability for comprehensive low-altitude communication research.

\textbf{LAIN-Radio3D}: The LAIN-Radio3D dataset is introduced to overcome the aforementioned limitations in vertical coverage and to better satisfy the requirements of low-altitude intelligent networks. Based on the open-source SionnaRT platform \cite{Sionna}, we conducted extensive RT simulations across these urban environments. While maintaining the same 2D building distributions as UrbanRadio3D, we employ different parameter settings.
First, the building heights are extended to a range from 15.8 m to 89.6 m, thereby introducing more complex non-line-of-sight (NLoS) propagation conditions. Second, to facilitate comprehensive research on low-altitude communications, the Rx height range is significantly elevated to 120\text{ m}. Third, unlike the fixed Tx configuration, LAIN-Radio3D incorporates variable Tx heights ranging from 7\text{ m} to 77\text{ m}. Finally, the carrier frequency is adjusted to 4.9\text{ GHz}, which is allocated as a dedicated frequency band for low-altitude communication networks. Overall, the extensive simulation process produces a large-scale dataset comprising 5.07 million RM slices. A detailed comparison of the key parameters between UrbanRadio3D and LAIN-Radio3D datasets is provided in Table~\ref{tab:dataset_comparison}.

\textbf{LAMBDA}: To further validate the generalization capability of the proposed framework in realistic urban environments, the San Francisco (SF) scenario from the recently released LAMBDA dataset is additionally adopted. Unlike UrbanRadio3D and LAIN-Radio3D, which are generated based on simplified urban representations, LAMBDA provides a high-quality digital twin constructed from real-world city geometries. The SF scenario contains detailed 3D building structures, realistic height distributions, and diverse material properties, including concrete, glass and brick, resulting in more complicated electromagnetic propagation conditions.

\subsubsection{Implementation Details}
The BDFlow-3DRM framework is implemented using the PyTorch library and trained on four NVIDIA L40 GPUs. The training procedure consists of two stages. In the first phase, the SVG encoder is trained to compress the RM into a compact semantic latent space. The latent representation consists of a 384-dimensional semantic feature extracted by the pre-trained DINOv3-s16p and a 24-dimensional embedding from the residual encoder, both with a spatial resolution of 16 $\times$ 16. The AdamW optimizer is employed with a batch size of 16 and a fixed learning rate of $1 \times 10^{-4}$. In the second stage, the 3D U-Net generative backbone and the proposed eta predictor are jointly optimized within the pre-trained latent space. The same batch size of 16 is adopted, and the learning rate follows a cosine decay schedule from $1 \times 10^{-4}$ down to $5 \times 10^{-5}$. For all subsequent experiments, we set the total number of available height layers $N = 20$ and the number of sampled height layers $D = 5$ to balance efficiency and performance.

\begin{table}[!t]
    \centering
    \caption{Dataset Parameters}
    \label{tab:dataset_comparison}
    \begin{tabular}{@{}ccc@{}}
        \toprule
        \textbf{Parameter} & \textbf{UrbanRadio3D} & \textbf{LAIN-Radio3D} \\
        \midrule
        Simulation Platform & WinProp  & SionnaRT \\
        Building Heights & $6.6 \sim 19.8\text{ m}$ & $15.8 \sim 89.6\text{ m}$ \\
        Tx Height & $1.5\text{ m}$ & 7, 12, 36, 47, 64, 77$\text{ m}$  \\
        Rx Height Range & $1\text{ m} \sim 20\text{ m}$ & $1\text{ m} \sim 120\text{ m}$ \\
        Carrier Frequency & $5.9\text{ GHz}$ & 4.9 \text{GHz} \\
        Power Max threshold & $-92.0$ \text{dBm} & $-60.0$ \text{dBm} \\
        Power Min threshold & $-169.0$ \text{dBm} & $-160.0$ \text{dBm} \\
        \bottomrule
    \end{tabular}
\end{table}

\subsection{Baselines and Evaluation Metrics}
We compare BDFlow-3DRM with the following representative baselines.

\textbf{3D U-Net}: An encoder--decoder baseline for 3D RM construction. Motivated by the effectiveness of U-Net-based architectures for RM prediction demonstrated by RadioUNet\cite{ref_RadioUNet}, we construct a 3D extension of the U-Net architecture to directly predict the complete 3D RM from environmental information and Tx configuration.

\textbf{3D-RadioDiff}\cite{ref_3DRadioDiff}: A diffusion-based framework that generates RMs separately for individual Rx heights under height-aware conditioning. Following the original design, the model is adapted to our experimental settings using identical input conditions and datasets. The complete 3D RM is obtained by independently generating and stacking all height layers.

\textbf{RadioDiff-3D}\cite{ref_RadioDiff3D}: A diffusion-based framework that directly generates 3D RM under the same environmental and Tx conditions as our work. It adopts a 3D U-Net backbone to model the full 3D RM volume.

To comprehensively evaluate the construction performance, we employ four quantitative metrics: NMSE, peak signal-to-noise ratio (PSNR), structural similarity index (SSIM), and power-level mean absolute error (Power-MAE). We evaluate all metrics on individual 2D horizontal slices before averaging them vertically. Let $\mathcal{R}_{k}, \hat{\mathcal{R}}_{k} \in \mathbb{R}^{L \times W}$ denote the ground truth and predicted 2D slices at height index $k \in \{1, \ldots, D\}$. 

NMSE captures the scale-independent relative error between the predicted and ground truth signals. For the $k$-th slice, it is defined as
\begin{equation}
    \text{NMSE}_k = \frac{\sum_{i=1}^{L} \sum_{j=1}^{W} (\mathcal{R}_k(i,j) - \hat{\mathcal{R}}_k(i,j))^2}{\sum_{i=1}^{L} \sum_{j=1}^{W} \mathcal{R}_k(i,j)^2}.
    \label{eq:nmse_k}
\end{equation}

PSNR measures the ratio between signal dynamic range and the construction error, expressed as
\begin{equation}
    \text{PSNR}_k = 10 \log_{10} \left( \frac{L_k^2 LW}{\sum_{i=1}^{L} \sum_{j=1}^{W} (\mathcal{R}_k(i,j) - \hat{\mathcal{R}}_k(i,j))^2} \right),
    \label{eq:psnr_k}
\end{equation}
where $L_k = \max(\mathcal{R}_k) - \min(\mathcal{R}_k)$ is the maximum dynamic range of the current data. 

SSIM evaluates structural similarity using luminance $l$, contrast $c$, and structure $s$, expressed as
\begin{align}
    l(\mathcal{R}_k, \hat{\mathcal{R}}_k) &= \frac{2\mu_{R}\mu_{\hat{R}} + C_1}{\mu_{R}^2 + \mu_{\hat{R}}^2 + C_1}, \\
    c(\mathcal{R}_k, \hat{\mathcal{R}}_k) &= \frac{2\sigma_{R}\sigma_{\hat{R}} + C_2}{\sigma_{R}^2 + \sigma_{\hat{R}}^2 + C_2}, \\
    s(\mathcal{R}_k, \hat{\mathcal{R}}_k) &= \frac{\sigma_{R\hat{R}} + C_3}{\sigma_{R}\sigma_{\hat{R}} + C_3},
\end{align}
where $\mu$ and $\sigma^2$ denote the mean and variance of the respective slices, and $\sigma_{R\hat{R}}$ is their covariance. The slice-wise index is $\text{SSIM}_k = l \cdot c \cdot s$. Constants $C_1, C_2, C_3$ avoid instability when denominators are close to zero.

To assess the construction accuracy in the original physical power domain (measured in dB), the normalized outputs are mapped back to their original signal power ranges. Let $\mathcal{R}_k(i,j)$ and $\hat{\mathcal{R}}_k(i,j)$ denote the normalized ground truth and predicted values, respectively. The corresponding physical powers $\mathcal{G}_k(i,j)$ and $\hat{\mathcal{G}}_k(i,j)$ are obtained via the de-normalization mappings
\begin{equation}
    \mathcal{G}_k = 
    \begin{cases} 
        \mathcal{R}_k \times 77 - 169, & \text{for UrbanRadio3D}, \\
        \mathcal{R}_k \times 100 - 160, & \text{for LAIN-Radio3D}.
    \end{cases}
    \label{eq:de_norm}
\end{equation}
The Power-MAE for the $k$-th slice is then calculated as the mean absolute deviation in the dB domain, expressed as
\begin{equation}
    \text{Power-MAE}_k = \frac{1}{LW} \sum_{i=1}^{L} \sum_{j=1}^{W} \left| \mathcal{G}_k(i,j) - \hat{\mathcal{G}}_k(i,j) \right|.
    \label{eq:gain_mae_k}
\end{equation}

\begin{figure}[!t]
\centering

\begin{minipage}{0.8\linewidth}
\centering

\begin{minipage}[c]{0.04\linewidth} 
    \centering
    \rotatebox{90}{\scriptsize Ground Truth} 
\end{minipage}\hspace{3pt}
\begin{minipage}[c]{0.30\linewidth}
    \centering
    \includegraphics[width=\linewidth]{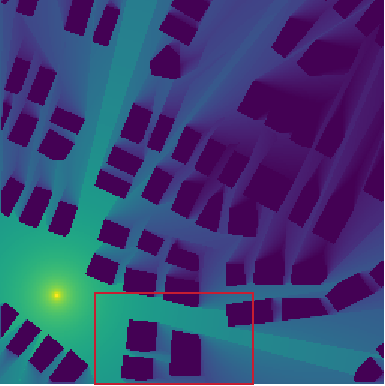}
\end{minipage}\hfill% 
\begin{minipage}[c]{0.30\linewidth}
    \centering
    \includegraphics[width=\linewidth]{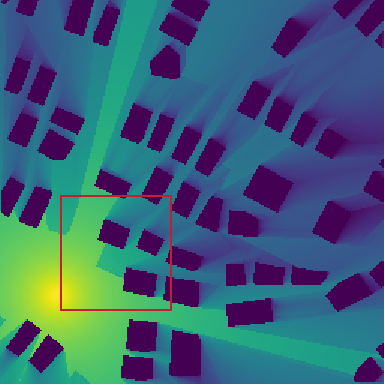}
\end{minipage}\hfill%
\begin{minipage}[c]{0.30\linewidth}
    \centering
    \includegraphics[width=\linewidth]{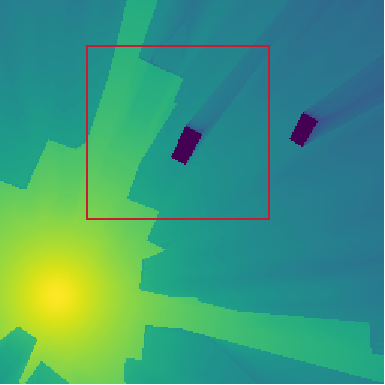}
\end{minipage}

\vspace{3pt}

\begin{minipage}[c]{0.04\linewidth}
    \centering
    \rotatebox{90}{\scriptsize 3D U-Net}
\end{minipage}\hspace{3pt}
\begin{minipage}[c]{0.30\linewidth}
    \centering
    \includegraphics[width=\linewidth]{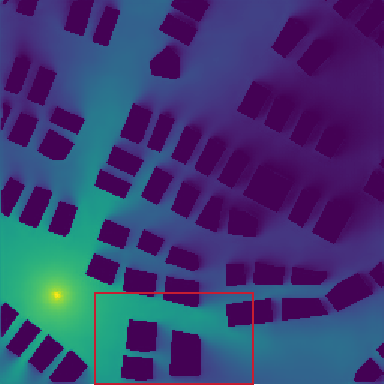}
\end{minipage}\hfill
\begin{minipage}[c]{0.30\linewidth}
    \centering
    \includegraphics[width=\linewidth]{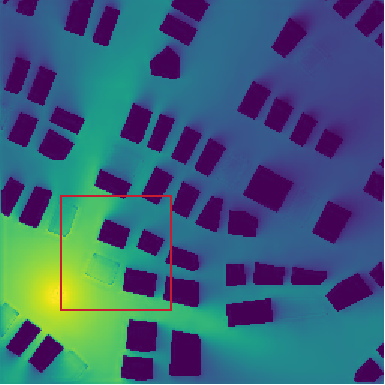}
\end{minipage}\hfill
\begin{minipage}[c]{0.30\linewidth}
    \centering
    \includegraphics[width=\linewidth]{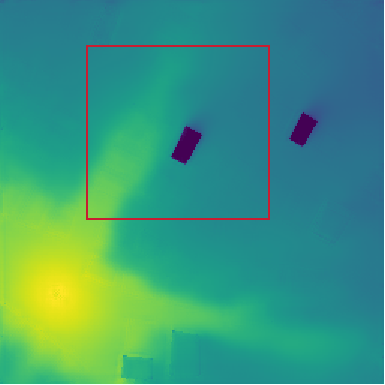}
\end{minipage}

\vspace{3pt}

\begin{minipage}[c]{0.04\linewidth}
    \centering
    \rotatebox{90}{\scriptsize 3D-RadioDiff}
\end{minipage}\hspace{3pt}
\begin{minipage}[c]{0.30\linewidth}
    \centering
    \includegraphics[width=\linewidth]{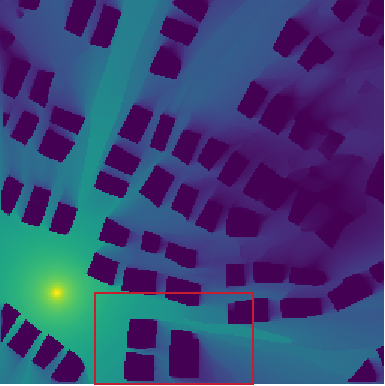}
\end{minipage}\hfill
\begin{minipage}[c]{0.30\linewidth}
    \centering
    \includegraphics[width=\linewidth]{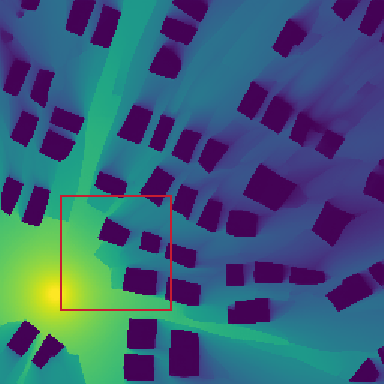}
\end{minipage}\hfill
\begin{minipage}[c]{0.30\linewidth}
    \centering
    \includegraphics[width=\linewidth]{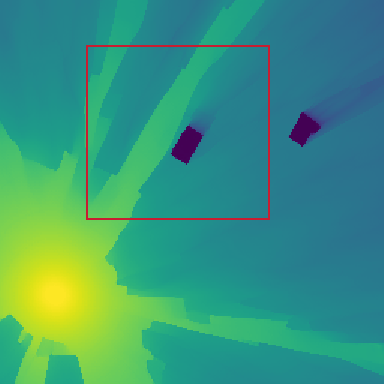}
\end{minipage}

\vspace{3pt}

\begin{minipage}[c]{0.04\linewidth}
    \centering
    \rotatebox{90}{\scriptsize RadioDiff-3D}
\end{minipage}\hspace{3pt}
\begin{minipage}[c]{0.30\linewidth}
    \centering
    \includegraphics[width=\linewidth]{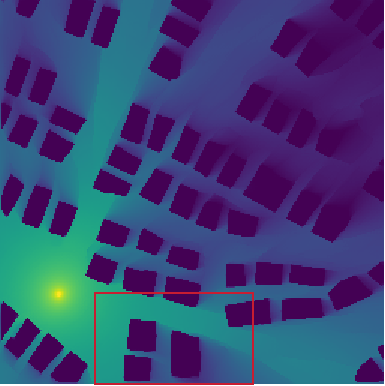}
\end{minipage}\hfill
\begin{minipage}[c]{0.30\linewidth}
    \centering
    \includegraphics[width=\linewidth]{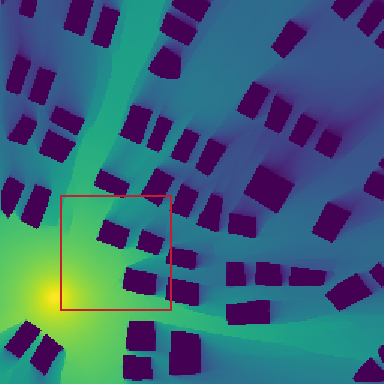}
\end{minipage}\hfill
\begin{minipage}[c]{0.30\linewidth}
    \centering
    \includegraphics[width=\linewidth]{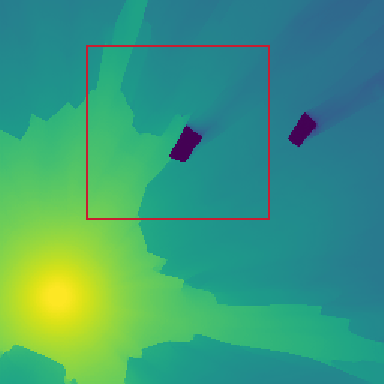}
\end{minipage}

\vspace{3pt}

\begin{minipage}[c]{0.04\linewidth}
    \centering
    \rotatebox{90}{\scriptsize BDFlow-3DRM}
\end{minipage}\hspace{3pt}
\begin{minipage}[c]{0.30\linewidth}
    \centering
    \includegraphics[width=\linewidth]{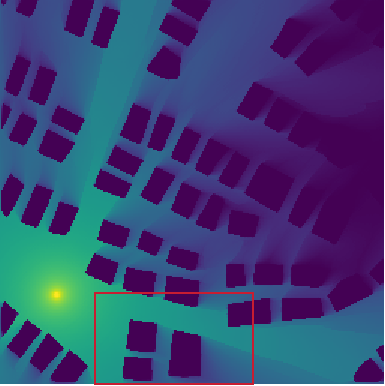}
\end{minipage}\hfill
\begin{minipage}[c]{0.30\linewidth}
    \centering
    \includegraphics[width=\linewidth]{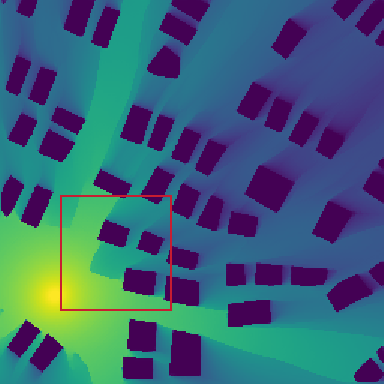}
\end{minipage}\hfill
\begin{minipage}[c]{0.30\linewidth}
    \centering
    \includegraphics[width=\linewidth]{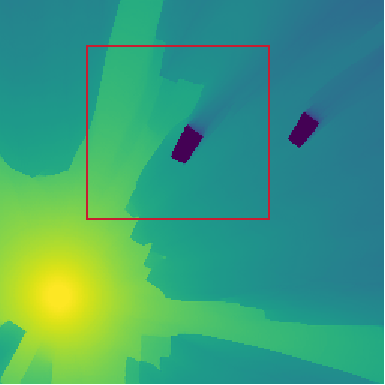}
\end{minipage}

\vspace{5pt}

\begin{minipage}{0.04\linewidth}
    ~ 
\end{minipage}\hspace{3pt}
\begin{minipage}{0.30\linewidth}
    \centering
    {\scriptsize H = 1 m}
\end{minipage}\hfill
\begin{minipage}{0.30\linewidth}
    \centering
    {\scriptsize H = 9 m}
\end{minipage}\hfill
\begin{minipage}{0.30\linewidth}
    \centering
    {\scriptsize H = 19 m}
\end{minipage}

\end{minipage}
\vspace{3pt}
\caption{Visual comparison of 3D RM construction results at different heights and with different models.} 
\label{fig:heatmap_comparison}
\end{figure}

\begin{table}[!t]
\centering
\setlength{\tabcolsep}{4pt}
\caption{Performance Comparison on UrbanRadio3D}
\label{tab:overall_performance}
\renewcommand{\arraystretch}{1}
\begin{tabular}{@{}ccccc@{}}
\toprule
\textbf{Methods} & \textbf{NMSE} $\downarrow$ & \textbf{PSNR(dB)} $\uparrow$ & \textbf{SSIM} $\uparrow$ & \textbf{Power-MAE(dB)} $\downarrow$ \\
\midrule
3D U-Net & 0.0549 & 22.638 & 0.7884 &  3.598 \\
3D-RadioDiff & 0.0521  &22.734  & 0.7950 & 3.452\\
RadioDiff-3D  & 0.0444 & 23.215 & 0.8067 & 3.116 \\
\textbf{BDFlow-3DRM} & \textbf{0.0317} & \textbf{24.656} & \textbf{0.8406} & \textbf{2.521} \\
\bottomrule
\end{tabular}
\end{table}

\begin{table}[!t]
    \centering
    \setlength{\tabcolsep}{1.5pt} 
    \caption{Ablation study on the Bi-Dynamical Block and Eta Predictor}
    \label{tab:ablation}
    \renewcommand{\arraystretch}{1}
    \begin{tabular}{@{}ccccc@{}}
        \toprule
        \textbf{Configuration} & \textbf{NMSE $\downarrow$} & \textbf{PSNR(dB) $\uparrow$} & \textbf{SSIM $\uparrow$} & \textbf{Power-MAE (dB) $\downarrow$} \\
        \midrule
        w/o correlation  & 0.0411 & 23.582 & 0.8090 & 2.959 \\
        static correlation  & 0.0388 & 23.856 & 0.8158 & 2.873 \\
        \textbf{adaptive correlation}  &\textbf{0.0317} & \textbf{24.656} & \textbf{0.8406} & \textbf{2.521} \\
        \bottomrule
    \end{tabular}
\end{table}

\subsection{Comparison With Baseline Methods}
\label{sec:benchmark_comparison}
The quantitative comparison between BDFlow-3DRM and three representative baseline methods is summarized in Table \ref{tab:overall_performance}. For a fair evaluation, all baseline methods are trained using complete RM observations across all Rx heights ranging from 1 m to 20 m. In contrast, BDFlow-3DRM is trained only with a sparse subset of height layers, including 1, 4, 7, 9, 11, 13, 15, 17, 19, and 20 m, while the evaluation is performed over the complete height range. Consequently, half of the Rx height layers are excluded from the training process, making the evaluation substantially more challenging for the proposed method.
Despite these sparse training observations, BDFlow-3DRM consistently achieves the best performance across all evaluation metrics. Specifically, it achieves an NMSE of 0.0317, corresponding to a 28.6$\%$ improvement compared with the RadioDiff-3D. Moreover, BDFlow-3DRM attains the lowest Power-MAE of 2.521 dB, indicating more accurate construction of signal power. The visual comparison presented in Fig.~\ref{fig:heatmap_comparison} further demonstrates that BDFlow-3DRM generates RMs with more faithful signal distributions and clearer structural details.

The performance gain can be attributed to the proposed combination of continuous height-aware representation learning and explicit inter-layer correlation modeling. By better capturing signal distribution correlations across height layers, the proposed design also enhances the reliability of prediction at unseen heights. As a result, BDFlow-3DRM maintains high construction accuracy even when a significant portion of Rx heights is excluded from training.

\subsection{Ablation Study}
We evaluate the effectiveness of the proposed inter-layer correlation mechanism by comparing the complete BDFlow-3DRM framework against two ablated configurations. The configuration without the bi-dynamical block sets the inter-layer correlation factor $\eta$ to 0, disabling the entire block and reverting to a standard architecture lacking vertical dependency. The static inter-layer correlation configuration enables this block but replaces the adaptive eta predictor with a fixed $\eta$ of 0.2, enforcing a uniform inter-layer coupling strength across all height layers. Finally, the adaptive inter-layer correlation setup represents our complete proposed method.

The quantitative results are summarized in Table~\ref{tab:ablation}. The configuration without inter-layer correlation exhibits the lowest construction accuracy. Introducing a fixed inter-layer coupling strategy improves all evaluation metrics, demonstrating that explicitly modeling vertical dependencies is essential for accurate 3D RM construction. However, the performance improvement remains limited because a fixed correlation coefficient cannot characterize the non-uniform dependencies among different height layers. In contrast, the proposed adaptive correlation mechanism achieves the best performance by enabling the eta predictor to dynamically adjust the inter-layer fusion strength based on local environmental features and geometric discrepancies. These results verify that both the bi-dynamical block and the adaptive correlation strategy significantly contribute to accurate 3D RM construction and improved construction coherence across height layers.

\begin{figure}[!t]
    \centering
    {
        \includegraphics[width=0.850\linewidth]{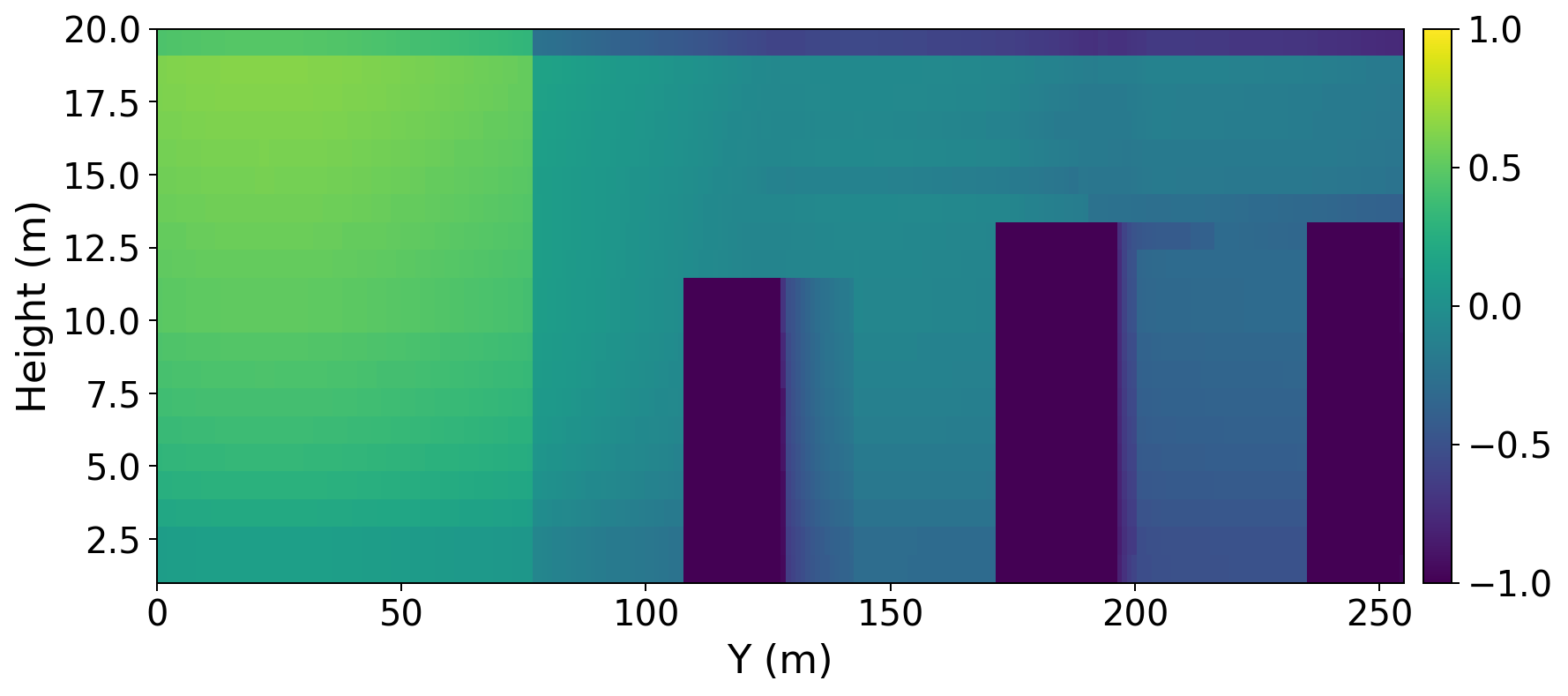}
    }
    \centerline{\scriptsize Ground Truth}

    {
        \includegraphics[width=0.850\linewidth]{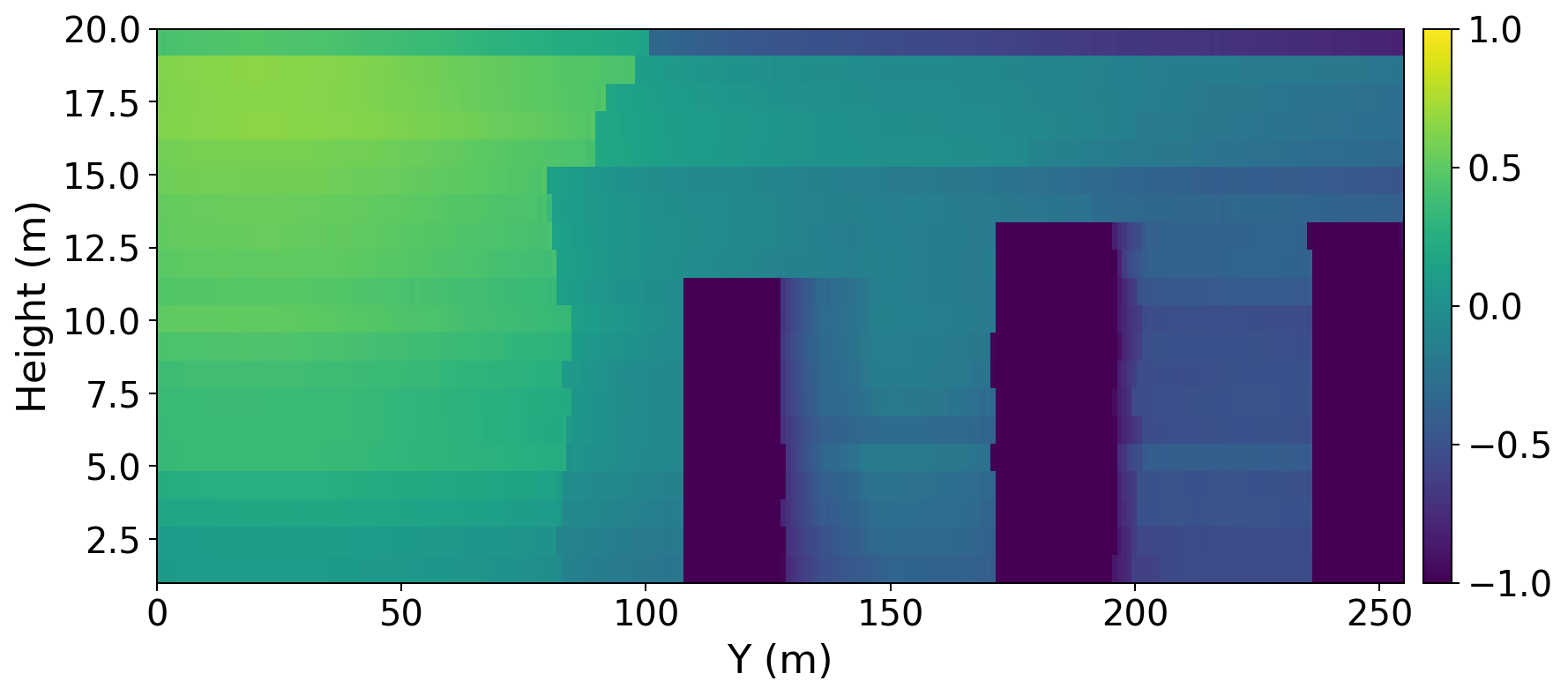}
    }
    \centerline{\scriptsize Adaptive correlation }

    {
        \includegraphics[width=0.850\linewidth]{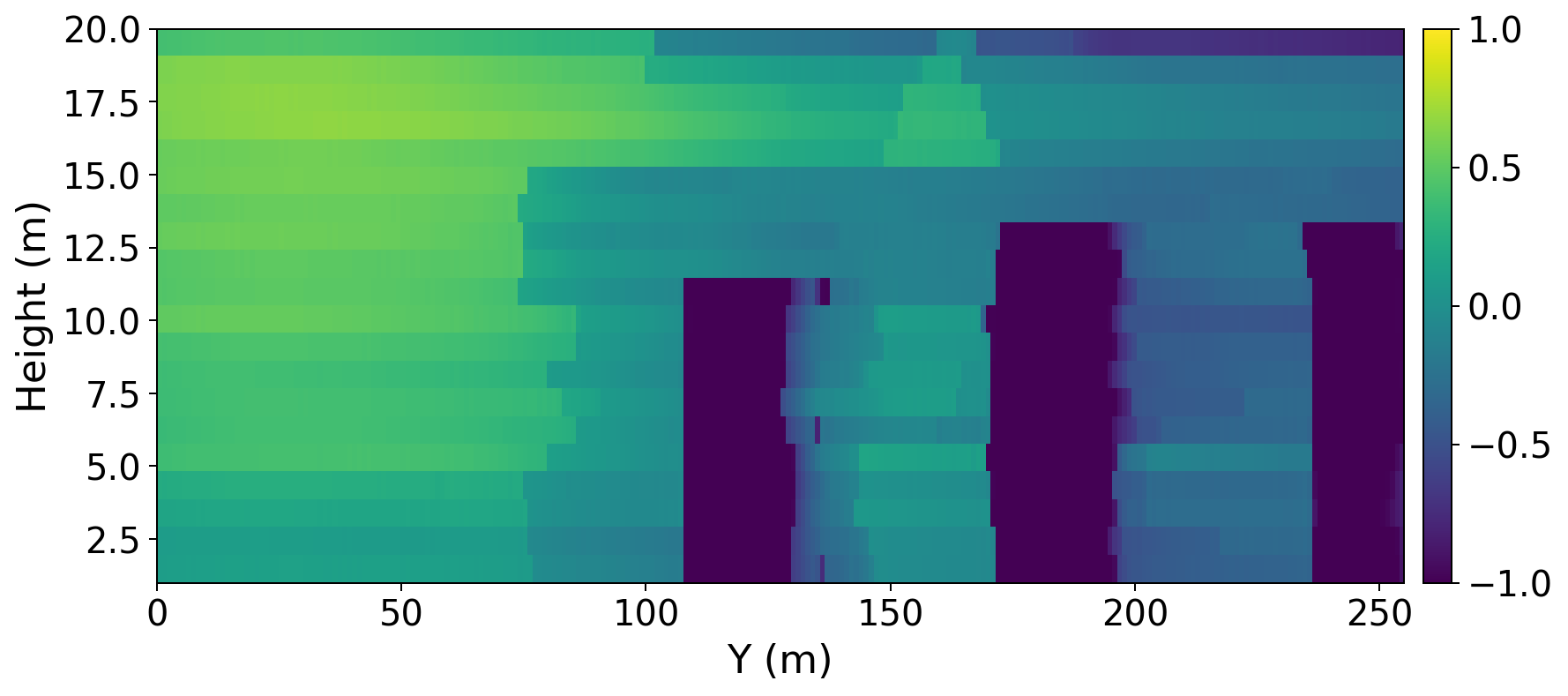}
    }
    \centerline{\scriptsize W/o correlation}
        
    \caption{Vertical profile comparison for a representative 3D RM slice at $x=64$. The horizontal axis is the $y$-direction spatial index in meters, while the vertical axis is the height in meters.}
    \label{fig:vertical_profile}
\end{figure}

\begin{figure}[!t]
    \centering
    \includegraphics[width=0.85\linewidth]{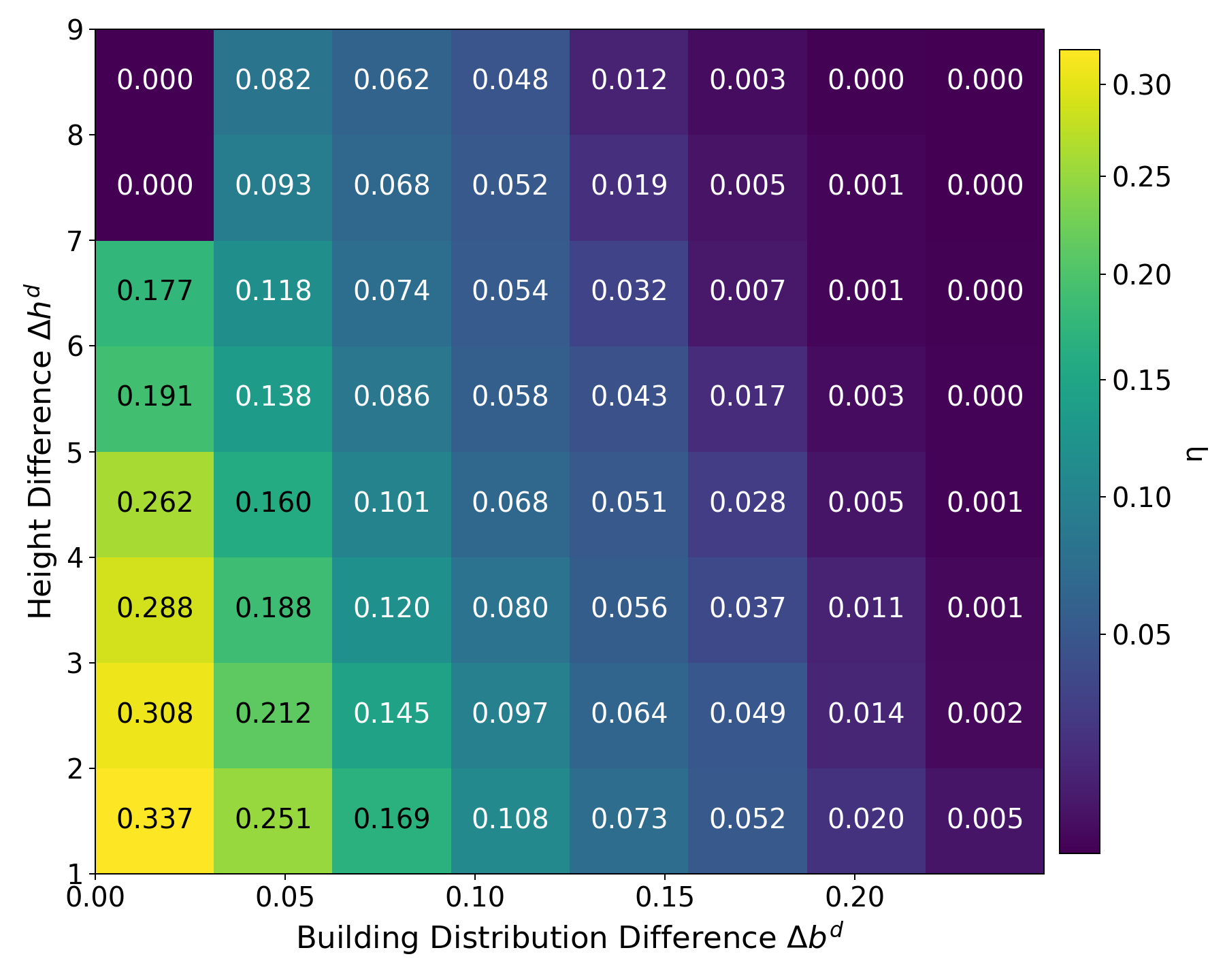}
    \caption{Heatmap of the predicted fusion coefficient $\eta$ under different inter-layer discrepancies. The horizontal axis is the building distribution difference $\Delta b^d$, and the vertical axis is the height difference $\Delta h^d$.}
    \label{fig:eta_heatmap}
    \vspace{-4pt}
\end{figure}

\subsection{Analysis of Height Coherence }
To evaluate the height coherence across different layers, we analyze the vertical profiles of the constructed 3D RM at a fixed horizontal location. As shown in Fig.~\ref{fig:vertical_profile}, the ground truth profile exhibits clear signal attenuation patterns caused by building blockages and smooth signal variations along the height dimension. 
The proposed BDFlow-3DRM with adaptive inter-layer correlation mechanism successfully reproduces these patterns. Although slight deviations from the ground truth are observed in regions with abrupt signal variations, the predicted profile preserves the overall height coherence of the signal distribution. 
In contrast, the variant without correlation exhibits more evident inconsistencies. In particular, the blocked regions are less well aligned with the ground truth, and the vertical transitions appear less coherent, especially around buildings. 
Overall, incorporating correlation enables more accurate recovery of building-induced attenuation patterns and improves the height coherence of the predicted signal distribution. This observation further verifies the effectiveness of the bi-dynamical block in preserving cross-layer coherence during 3D RM construction.

\subsection{Behavior of the Eta Predictor}
A detailed analysis of the learned fusion coefficients provides additional insight into the behavior of the proposed adaptive correlation mechanism. Since the eta predictor is designed to regulate the fusion strength between neighboring heights according to their relative differences, we examine how the predicted correlation coefficients vary with two factors: the inter-layer height difference $\Delta h^d$ and the building distribution difference $\Delta b^d$.

As shown in Fig.~\ref{fig:eta_heatmap}, the learned correlation coefficient $\eta$ exhibits a clear dependency on both $\Delta h^d$ and $\Delta b^d$, showing a monotonic decreasing trend as the discrepancy score increases. The maximum value appears in the low-score region $(\Delta h^d, \Delta b^d) \approx (1, 0)$ with $\eta = 0.337$, where neighboring heights tend to share more similar propagation patterns. As the discrepancy score increases, $\eta$ decreases continuously toward near-zero values (e.g., for $\Delta h^d \geq 8$ and $\Delta b^d \geq 0.20$, $\eta \in [0, 0.001]$). This trend is consistent with our design intuition that neighboring heights with smaller cross-layer differences are more likely to benefit from stronger coupling. Combined with the ablation results in Table~\ref{tab:ablation}, these observations suggest that the adaptive correlation mechanism benefits from assigning fusion weights in a manner that is consistent with local cross-layer differences, thereby contributing to more height-coherent 3D RM construction. 

\begin{figure*}[!t]
    \centering
    
    \noindent\makebox[\textwidth][c]{%
    \begin{minipage}[c]{0.04\textwidth}
        \centering
        \rotatebox{90}{\scriptsize Ground Truth}
    \end{minipage}\hspace{3pt}%
    \begin{minipage}[c]{0.8\textwidth}
        \centering

        \begin{minipage}{0.16\linewidth}
            \centering
            \includegraphics[width=\linewidth]{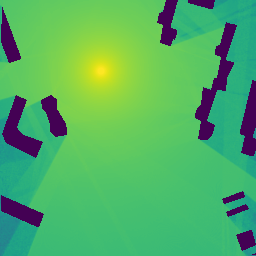}\\
            \scriptsize (101,185,12)$|$H=8
        \end{minipage}\hfill
        \begin{minipage}{0.16\linewidth}
            \centering
            \includegraphics[width=\linewidth]{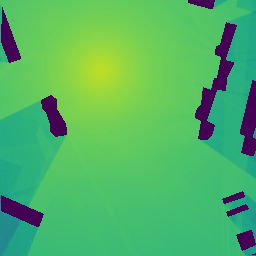}\\
            \scriptsize (101,185,12)$|$H=28
        \end{minipage}\hfill
        \begin{minipage}{0.16\linewidth}
            \centering
            \includegraphics[width=\linewidth]{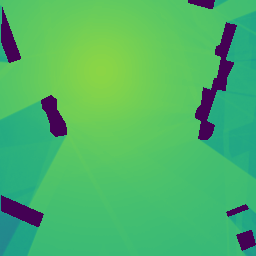}\\
            \scriptsize (101,185,12)$|$H=48
        \end{minipage}\hfill
        \begin{minipage}{0.16\linewidth}
            \centering
            \includegraphics[width=\linewidth]{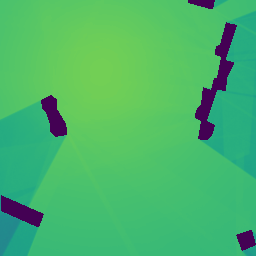}\\
            \scriptsize (101,185,12)$|$H=68
        \end{minipage}\hfill
        \begin{minipage}{0.16\linewidth}
            \centering
            \includegraphics[width=\linewidth]{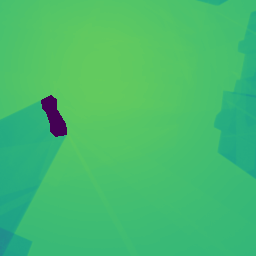}\\
            \scriptsize (101,185,12)$|$H=88
        \end{minipage}\hfill
        \begin{minipage}{0.16\linewidth}
            \centering
            \includegraphics[width=\linewidth]{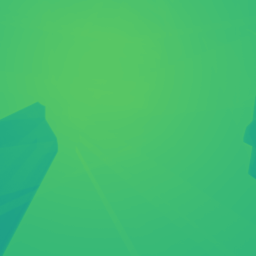}\\
            \scriptsize (101,185,12)$|$H=108
        \end{minipage}

    \end{minipage}%
    }

    \vspace{3pt}

    \noindent\makebox[\textwidth][c]{%
    \begin{minipage}[c]{0.04\textwidth}
        \centering
        \rotatebox{90}{\scriptsize BDFlow-3DRM}
    \end{minipage}\hspace{3pt}%
    \begin{minipage}[c]{0.8\textwidth}
        \centering

        \begin{minipage}{0.16\linewidth}
            \centering
            \includegraphics[width=\linewidth]{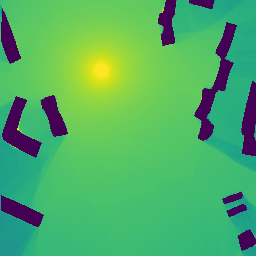}\\
            \scriptsize (101,185,12)$|$H=8
        \end{minipage}\hfill
        \begin{minipage}{0.16\linewidth}
            \centering
            \includegraphics[width=\linewidth]{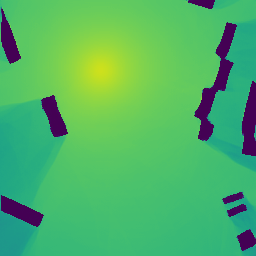}\\
            \scriptsize (101,185,12)$|$H=28
        \end{minipage}\hfill
        \begin{minipage}{0.16\linewidth}
            \centering
            \includegraphics[width=\linewidth]{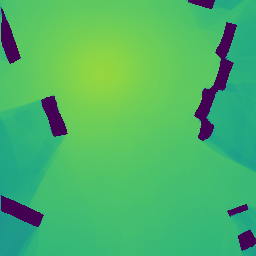}\\
            \scriptsize (101,185,12)$|$H=48
        \end{minipage}\hfill
        \begin{minipage}{0.16\linewidth}
            \centering
            \includegraphics[width=\linewidth]{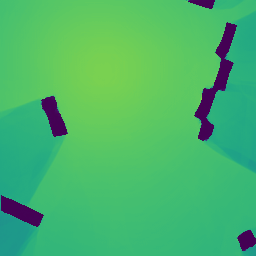}\\
            \scriptsize (101,185,12)$|$H=68
        \end{minipage}\hfill
        \begin{minipage}{0.16\linewidth}
            \centering
            \includegraphics[width=\linewidth]{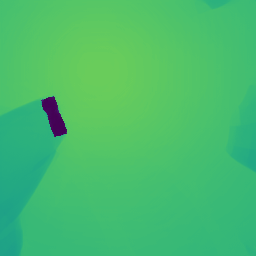}\\
            \scriptsize (101,185,12)$|$H=88
        \end{minipage}\hfill
        \begin{minipage}{0.16\linewidth}
            \centering
            \includegraphics[width=\linewidth]{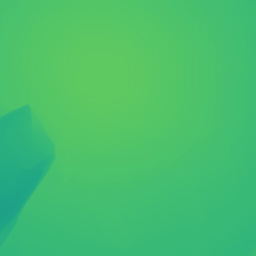}\\
            \scriptsize (101,185,12)$|$H=108
        \end{minipage}

    \end{minipage}%
    }
    \vspace{3pt}
    \caption{Visualization of zero-shot generalization on unseen Rx heights across the urban scenario. The annotation $(\cdot)$ denotes the 3D coordinates $(x, y, z)$ of the Tx in meters, and $H$ indicates the specific height of the Rx plane in meters.}
    \label{fig:rx_generalization}

\end{figure*}

\subsection{Generalization Test}
The generalization capability of BDFlow-3DRM is comprehensively evaluated on the LAIN-Radio3D dataset under unseen Rx and Tx height configurations, as well as on realistic urban environments from the LAMBDA dataset. In Tables~\ref{tab:rx_generalization} and~\ref{tab:tx_generalization}, Training Avg. denotes the average performance over the height configurations that are included in training.

\subsubsection{Rx Height Generalization}
To assess the generalization capability across different Rx heights, the model is trained using RMs corresponding to a sparse set of Rx heights, including 1 m, heights from 5 to 70 m with an interval of 5 m, and heights from 80 to 120 m with an interval of 10 m. During training, the Tx heights are set to 12 m, 36 m, and 64 m. The zero-shot inference performance is evaluated at unseen Rx heights of 8 m, 28 m, 48 m, 68 m, 88 m, and 108 m within the training height range, as well as extrapolated heights of 150 m and 200 m. Note that the results at 150~m and 200~m are obtained from additional RT-simulated test samples generated under the same setup as LAIN-Radio3D, and these samples are used only for evaluation. As shown in Table~\ref{tab:rx_generalization} and Fig.~\ref{fig:rx_generalization}, BDFlow-3DRM maintains high accuracy across both interpolated and extrapolated height layers, demonstrating the generalization capability even beyond the training height range. The performance generally improves at higher layers, where line-of-sight propagation becomes dominant, whereas lower layers remain more challenging due to complicated NLoS interactions caused by building blockages.

\begin{table}[!t]
    \centering
    \setlength{\tabcolsep}{4pt}
    \caption{Zero-Shot Generalization on Rx Heights}
    \label{tab:rx_generalization}
    \begin{tabular}{@{}ccccc@{}}
        \toprule
        \textbf{Rx Height} & \textbf{NMSE $\downarrow$} & \textbf{PSNR(dB) $\uparrow$} & \textbf{SSIM $\uparrow$} & \textbf{Power-MAE (dB) $\downarrow$} \\
        \midrule
        Training Avg.  & 0.0134 & 25.677 & 0.8448 & 2.985 \\
        \midrule
         $8\text{ m}$ & 0.0229 & 22.259 & 0.7938 & 3.468 \\
         $28\text{ m}$ & 0.0183 & 22.897 & 0.8105 & 3.293 \\
         $48\text{ m}$ & 0.0139 & 23.837 & 0.8239 & 3.250 \\
         $68\text{ m}$ & 0.0096 & 25.871 & 0.8634 & 2.829 \\
         $88\text{ m}$ & 0.0066 & 29.193 & 0.8927 & 2.711 \\
         $108\text{ m}$ & 0.0041 & 32.750 & 0.9280 & 2.014 \\
         $150\text{ m}$ & 0.0043 & 31.561 & 0.9472 & 2.581 \\
         $200\text{ m}$ & 0.0065 & 28.698 & 0.9447 & 3.879 \\
        \bottomrule
    \end{tabular}
\end{table}

\begin{table}[!t]
    \centering
    \setlength{\tabcolsep}{4pt}
    \caption{Zero-Shot Generalization on Tx Heights}
    \label{tab:tx_generalization}
    \begin{tabular}{@{}ccccc@{}}
        \toprule
        \textbf{Tx Height} & \textbf{NMSE $\downarrow$} & \textbf{PSNR(dB) $\uparrow$} & \textbf{SSIM $\uparrow$} & \textbf{Power-MAE (dB) $\downarrow$} \\
        \midrule
        Training Avg. & 0.0134 & 25.677 & 0.8448 & 2.985 \\
        \midrule
        $7\text{ m}$ & 0.0141 & 24.893 & 0.8248 & 3.375 \\
        $47\text{ m}$ & 0.0130 & 25.485 & 0.8531 & 3.076 \\
        $77\text{ m}$ & 0.0123 & 26.766 & 0.8779 & 2.773 \\
        \bottomrule
    \end{tabular}
\end{table}

\subsubsection{Tx Height Generalization} 
\begin{figure*}[!t]
    \centering

    \noindent\makebox[\textwidth][c]{%
    \begin{minipage}[c]{0.04\textwidth}
        \centering
        \rotatebox{90}{\scriptsize Ground Truth}
    \end{minipage}\hspace{3pt}%
    \begin{minipage}[c]{0.8\textwidth}
        \centering
        \begin{minipage}{0.16\linewidth}
            \centering
            \includegraphics[width=\linewidth]{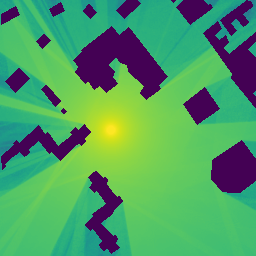}\\
            \scriptsize (111,126,7)$|$H=10
        \end{minipage}\hfill
        \begin{minipage}{0.16\linewidth}
            \centering
            \includegraphics[width=\linewidth]{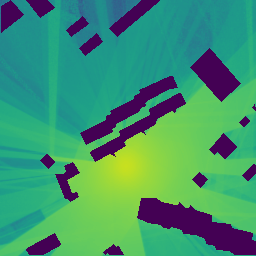}\\
            \scriptsize (126,90,7)$|$H=20
        \end{minipage}\hfill
        \begin{minipage}{0.16\linewidth}
            \centering
            \includegraphics[width=\linewidth]{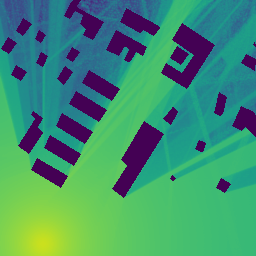}\\
            \scriptsize (43,12,47)$|$H=35
        \end{minipage}\hfill
        \begin{minipage}{0.16\linewidth}
            \centering
            \includegraphics[width=\linewidth]{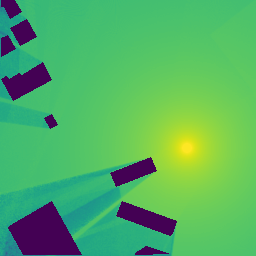}\\
            \scriptsize (187,108,47)$|$H=45
        \end{minipage}\hfill
        \begin{minipage}{0.16\linewidth}
            \centering
            \includegraphics[width=\linewidth]{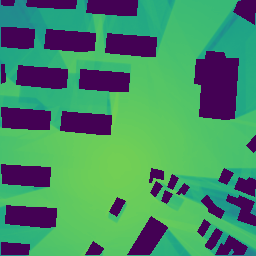}\\
            \scriptsize (117,87,77)$|$H=20
        \end{minipage}\hfill
        \begin{minipage}{0.16\linewidth}
            \centering
            \includegraphics[width=\linewidth]{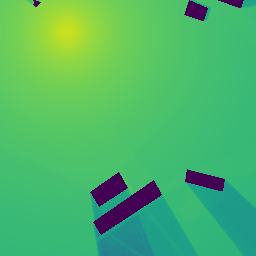}\\
            \scriptsize (67,224,77)$|$H=65
        \end{minipage}

    \end{minipage}%
    }

    \vspace{3pt}

    \noindent\makebox[\textwidth][c]{%
    \begin{minipage}[c]{0.04\textwidth}
        \centering
        \rotatebox{90}{\scriptsize BDFlow-3DRM}
    \end{minipage}\hspace{3pt}%
    \begin{minipage}[c]{0.8\textwidth}
        \centering
        \begin{minipage}{0.16\linewidth}
            \centering
            \includegraphics[width=\linewidth]{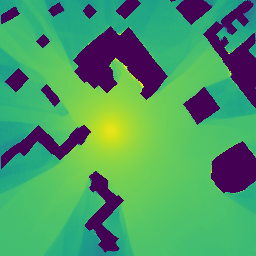}\\
            \scriptsize (111,126,7)$|$H=10
        \end{minipage}\hfill
        \begin{minipage}{0.16\linewidth}
            \centering
            \includegraphics[width=\linewidth]{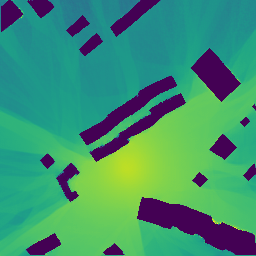}\\
            \scriptsize (126,90,7)$|$H=20
        \end{minipage}\hfill
        \begin{minipage}{0.16\linewidth}
            \centering
            \includegraphics[width=\linewidth]{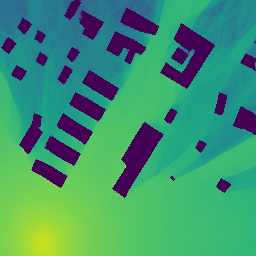}\\
            \scriptsize (43,12,47)$|$H=35
        \end{minipage}\hfill
        \begin{minipage}{0.16\linewidth}
            \centering
            \includegraphics[width=\linewidth]{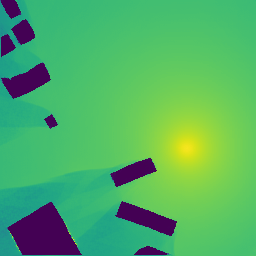}\\
            \scriptsize (187,108,47)$|$H=45
        \end{minipage}\hfill
        \begin{minipage}{0.16\linewidth}
            \centering
            \includegraphics[width=\linewidth]{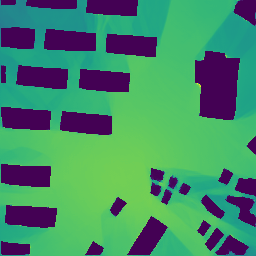}\\
            \scriptsize (117,87,77)$|$H=20
        \end{minipage}\hfill
        \begin{minipage}{0.16\linewidth}
            \centering
            \includegraphics[width=\linewidth]{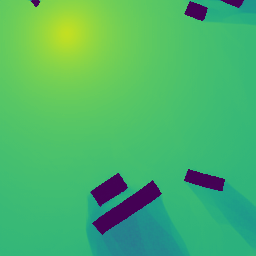}\\
            \scriptsize (67,224,77)$|$H=65
        \end{minipage}

    \end{minipage}%
    }
    \vspace{3pt}
    \caption{Visualization of zero-shot generalization on unseen Tx heights. From left to right, the paired columns represent zero-shot inference for Tx deployed at heights of 7\,m (columns 1--2), 47\,m (columns 3--4), and 77\,m (columns 5--6), respectively. The annotation $(\cdot)$ denotes the Tx 3D coordinates, and H indicates the Rx plane height in meters, consistent with Fig.~\ref{fig:rx_generalization}.}
    \label{fig:tx_generalization}
\end{figure*}

In addition to the variation in Rx heights, the ability to accurately predict RMs under unseen Tx heights is also essential. For this evaluation, we train the model with Tx deployments at 12\text{ m}, 36\text{ m} and 64\text{ m}, and test the model on unseen heights of 7\text{ m}, 47\text{ m} and 77\text{ m}, with Rx heights covering the 1-120 m range. As depicted in Table~\ref{tab:tx_generalization} and the visualization in Fig.~\ref{fig:tx_generalization}, BDFlow-3DRM delivers stable performance on entirely new Tx configurations. For instance, testing at a Tx height of 47 m yields an NMSE of 0.0130 and an SSIM of 0.8531, which are comparable to the performance achieved within the training height range. These results validate the proposed dual height encoding mechanism, enabling robust generalization across diverse transceiver height combinations for flexible 3D RM construction.

\begin{figure}[!t]
    \centering
    {
        \includegraphics[width=0.65\linewidth]{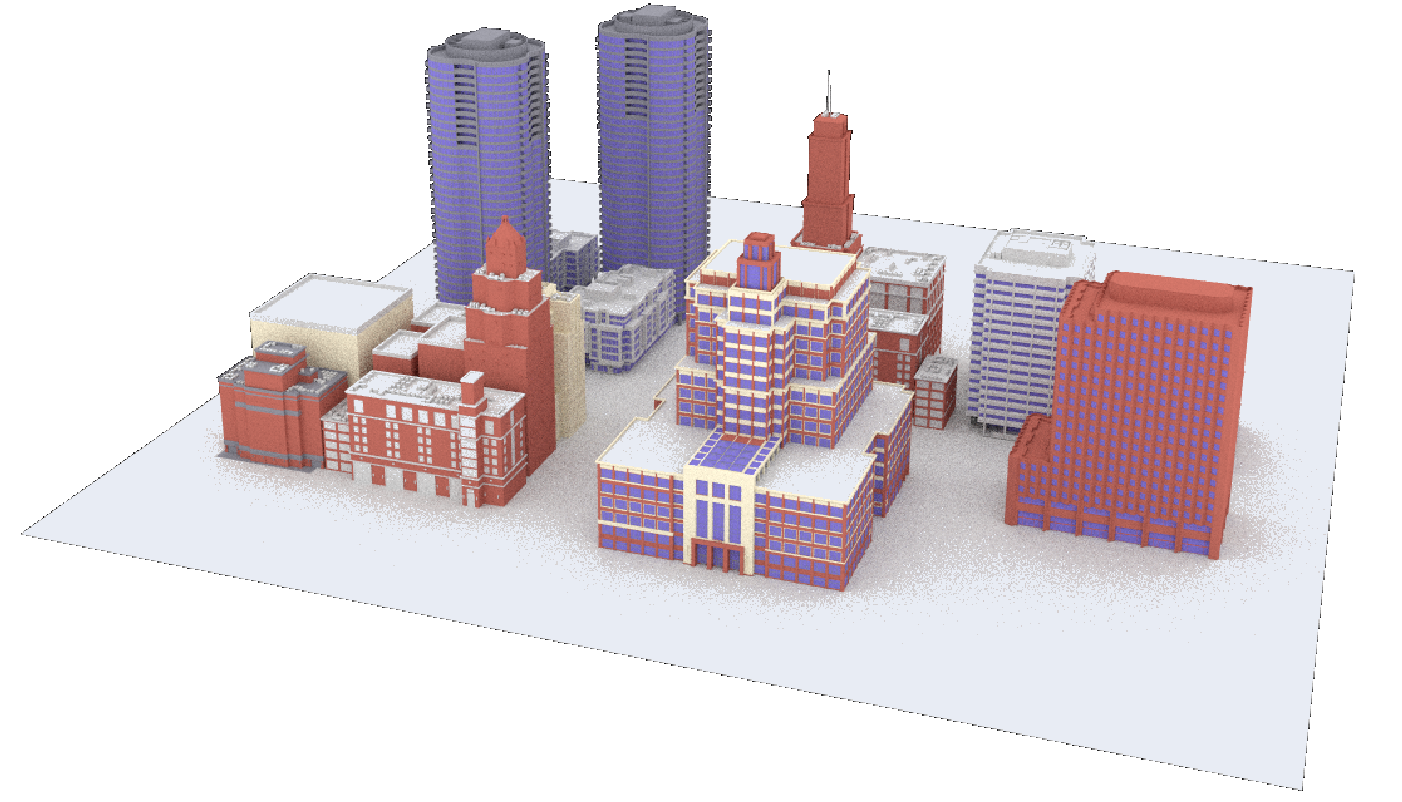} 
    }
    \centerline{\scriptsize (a) Realistic SF Building Environment}
    {
        \includegraphics[width=0.65\linewidth]{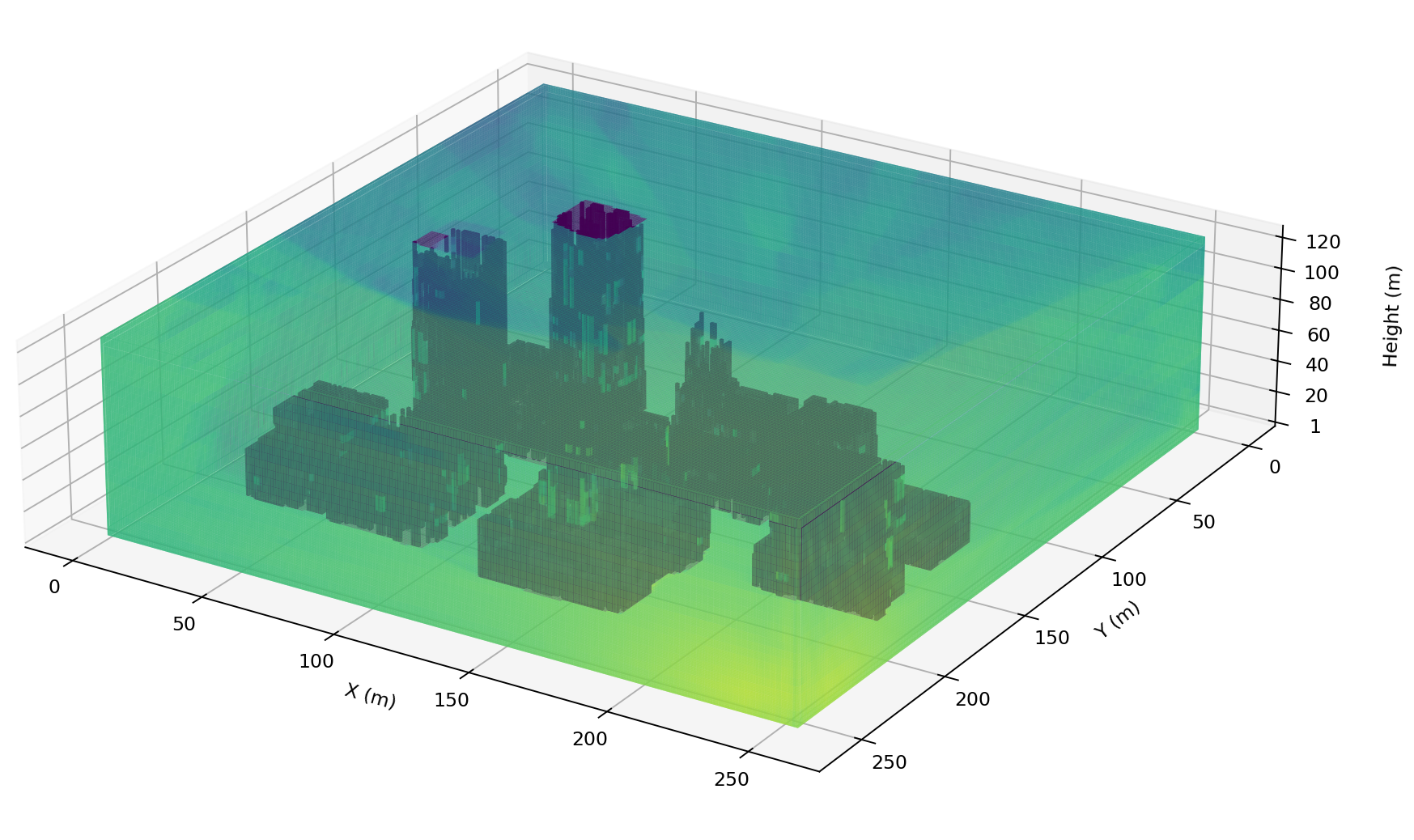} 
    }
    \centerline{\scriptsize (b) Predicted 3D RM}
    \caption{Visualization of the SF evaluation scenario: (a) the fine-grained 3D building environment, and (b) the full 3D RM predicted by BDFlow-3DRM without fine-tuning.}
    \label{fig:sf_visualization}
\end{figure}

\subsubsection{Generalization to Realistic Environments} 
\begin{figure}[t]
    \centering
    \includegraphics[width=0.8\linewidth]{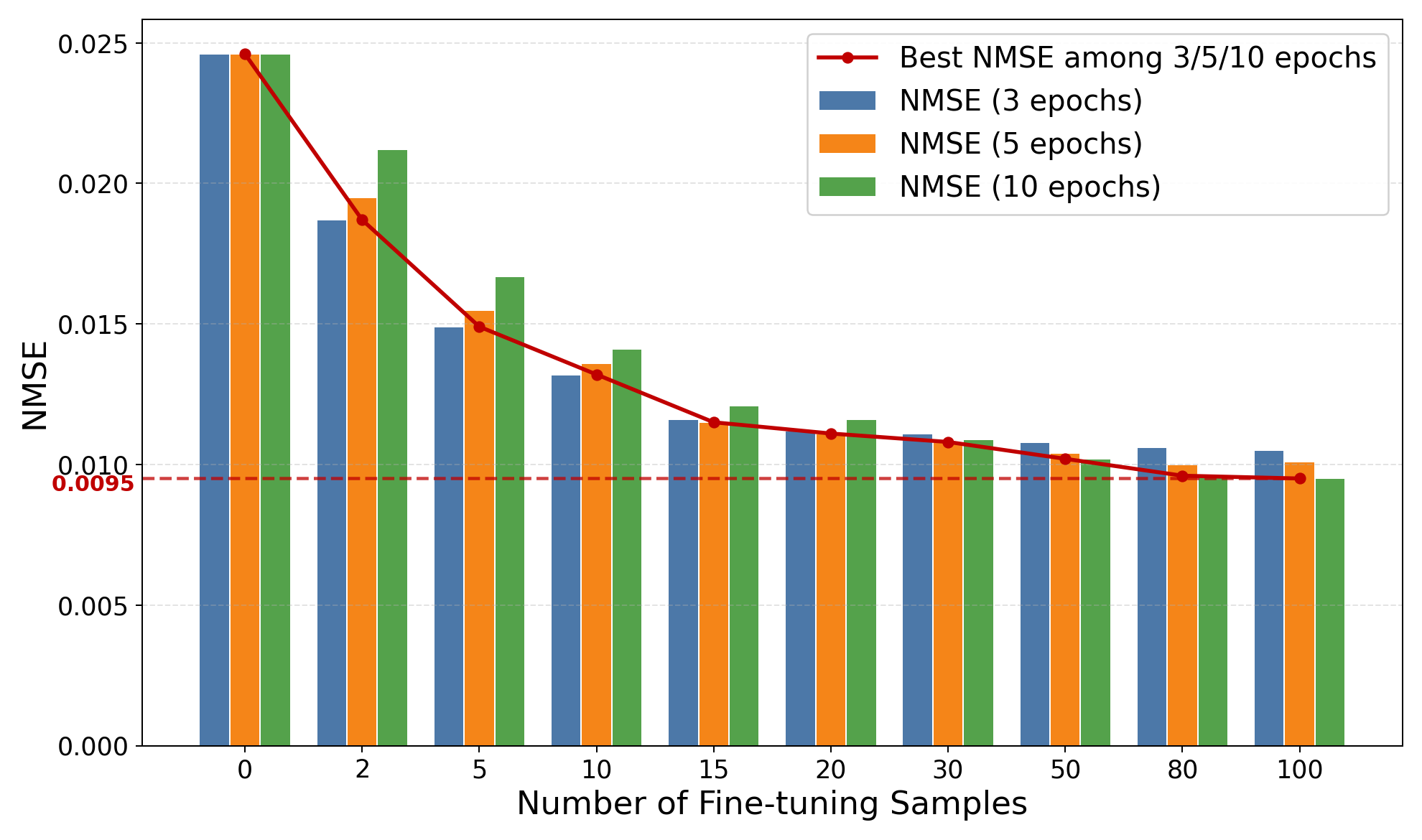} 
    \caption{Few-shot generalization performance of BDFlow-3DRM in the SF environment under different numbers of fine-tuning samples. The blue, orange, and green bars correspond to the NMSE achieved after 3, 5, and 10 fine-tuning epochs, respectively, while the red curve shows the best NMSE among these three settings.}
    \label{fig:finetune}
\end{figure}

To evaluate the cross-dataset transferability in realistic urban environments, we conduct zero-shot transfer experiments on the LAMBDA dataset. The evaluation involves four SF sub-scenarios with Tx heights of 7 m, 12 m, 36 m, 47 m, 64 m and 77 m, and an Rx height range from 1 m to 120 m. A representative environment layout and the corresponding predicted complete 3D RM are illustrated in Fig.~\ref{fig:sf_visualization}. Despite the unknown material properties and irregular real-world geometric structures, BDFlow-3DRM demonstrates robust zero-shot transferability, achieving an NMSE of 0.0246, an SSIM of 0.8021, and a Power-MAE of 4.19 dB without any fine-tuning. Furthermore, as shown in Fig.~\ref{fig:finetune}, the model exhibits strong few-shot adaptability: incorporating only 15--20 fine-tuning samples reduces the NMSE by more than 50\%, while performance gradually converges to a lower bound of approximately 0.0095 as the number of fine-tuning samples increases. These results suggest promising transferability and deployment potential of BDFlow-3DRM in realistic low-altitude urban environments.

\subsection{Efficiency and Complexity Analysis}
To evaluate computational efficiency, we compare the inference complexity of BDFlow-3DRM with that of RadioDiff-3D. The proposed method exhibits an advantage in reducing floating-point operations (FLOPs). Specifically, while RadioDiff-3D relies on a 1000-step Markov chain that accumulates approximately 2850.28\,TFLOPs during inference, BDFlow-3DRM formulates the generation process as a deterministic probability flow and employs the first-order Euler method for ODE integration. Consequently, high-quality 3D RM construction can be achieved using only 1 to 5 integration steps. This formulation reduces the total inference complexity to a range of 7.22 to 15.65\,TFLOPs, representing an over 180-fold reduction in computational cost.

We further investigate the impact of ODE integration steps on construction performance. As presented in Table~\ref{tab:steps_tradeoff}, BDFlow-3DRM achieves the best construction performance at 5 steps, obtaining an NMSE of 0.0317 and an SSIM of 0.8406 with an inference latency of only 0.283 seconds. Even when the number of integration steps is reduced to 1, the model maintains competitive performance, achieving an NMSE of 0.0323 while requiring only 0.107 seconds. Conversely, increasing the integration steps to 10, 25, or 50 leads to linear latency growth and a slight performance degradation, which is likely caused by the accumulation of numerical approximation errors during the iterative ODE integration process. These results demonstrate that BDFlow-3DRM can efficiently generate high-quality 3D RM with very few integration steps, enabling rapid inference without compromising accuracy.

\begin{table}[!t]
    \centering
    \caption{Impact of ODE Integration Steps on Performance}
    \label{tab:steps_tradeoff}
    \begin{tabular}{ccccc}
        \toprule
        \textbf{Steps} & \textbf{Latency (s) $\downarrow$} & \textbf{NMSE $\downarrow$} & \textbf{PSNR(dB) $\uparrow$} & \textbf{SSIM $\uparrow$} \\
        \midrule
        1  & \textbf{0.107} & 0.0323 & 24.511 & 0.8370 \\
        5  & 0.283 &\textbf{0.0317} & \textbf{24.656} & \textbf{0.8406} \\
        10 & 0.472 & 0.0335 & 24.399 & 0.8326 \\
        25 & 1.143 & 0.0350 & 24.210 & 0.8261 \\
        50 & 2.189 & 0.0359 & 24.105 & 0.8225 \\
        \bottomrule
    \end{tabular}
\end{table}

\section{Conclusion}
In this paper, we developed BDFlow-3DRM, a bi-dynamical flow matching framework for 3D RM construction that addresses a key limitation of existing methods, namely, insufficient modeling of the height dimension. Specifically, the 3D RM construction problem is formulated as a deterministic probability flow in a semantic latent space. By integrating continuous height-aware representation learning with bidirectional inter-layer dependency modeling, BDFlow-3DRM improves height generalization and cross-height construction coherence. Extensive experiments on the UrbanRadio3D, LAIN-Radio3D, and LAMBDA datasets confirm the effectiveness of BDFlow-3DRM. Compared with diffusion-based baselines, BDFlow-3DRM reduces the NMSE by 28.6\% and attains a 180-fold reduction in inference complexity. More importantly, with only 20 training Rx-height layers, it maintains accurate prediction over a wide continuous Rx-height range from 1 to 120\,m under variable Tx heights. Together with its robust zero-shot and few-shot transferability to realistic urban environments, these results highlight the practical value of BDFlow-3DRM for large-scale 3D RM construction in complex low-altitude scenarios. Future work will focus on extending the proposed framework beyond received signal strength toward comprehensive 3D CKM construction involving multiple channel metrics, including AoA/AoD and propagation delays. In addition, integrating ultra-sparse measurements into the framework to adapt the model to dynamic wireless environments represents a promising direction for bridging the gap between simulation-based training and practical deployment.

\bibliographystyle{IEEEtran}
\bibliography{ref}

\vfill

\end{document}